\documentclass[twocolumn,aps,superscriptaddress,nolongbibliography,pra,10pt]{revtex4-2}
\usepackage{graphicx}
\usepackage{amsmath}
\usepackage{amssymb}
\usepackage{bm}
\usepackage{hyperref}
\usepackage[dvipsnames]{xcolor}
\usepackage{tabularx}
\usepackage{placeins}

\usepackage[normalem]{ulem}

\begin{document}

\title{Mapping the parameter space of double microwave shielding}

\author{Hubert J. Jóźwiak}
\email{hubert.jozwiak@ru.nl}
\affiliation{Institute for Molecules and Materials, Radboud University, Nijmegen, The Netherlands}
\author{Ian Stevenson}
\affiliation{Department of Physics, Columbia University, New York, New York 10027, USA}
\affiliation{Department of Physics, University of Virginia, Charlottesville, Virginia 29904, USA}
\author{Sebastian Will}
\affiliation{Department of Physics, Columbia University, New York, New York 10027, USA}
\author{Tijs Karman}
\email{tkarman@science.ru.nl}
\affiliation{Institute for Molecules and Materials, Radboud University, Nijmegen, The Netherlands}

\begin{abstract}
    Double microwave shielding employs $\sigma^{+}$- and $\pi$-polarized microwave fields, tuned close to the lowest rotational transition, to engineer a long-range repulsive barrier between polar molecules. By preventing molecules from reaching the short range, this technique suppresses detrimental two-body losses and recently enabled the realization of molecular Bose-Einstein condensates and self-bound droplets. Yet, the optimal operating regimes of the shielding mechanism remain largely unexplored. Here, by leveraging the underlying universality of the scattering problem, we systematically map the four-dimensional microwave parameter space---spanned by the detunings and intensities of the two fields---to identify configurations that maximize both shielding efficiency and interaction tunability. We define optimal operating regimes as configurations that are strictly free of field-linked bound states while sufficiently suppressing two-body losses to exceed typical lifetimes of ultracold samples. In these regimes, we evaluate the elastic-to-inelastic collision ratios required for efficient evaporative cooling and explore the accessible tuning range of the effective dipolar interactions. Finally, to identify the best platforms for future quantum simulation experiments, we conduct a global survey of candidate molecular species under realistic field constraints. We identify heavy, strongly dipolar molecules as the most promising candidates, demonstrating that they can achieve extreme loss suppression{---reaching two-body loss rates as low as $10^{-18}$~cm$^{3}$/s---}alongside robust interaction tunability using only moderate field strengths.
\end{abstract}

\maketitle

\section{Introduction}

Ultracold polar molecules have emerged as versatile systems with applications ranging from precision tests of fundamental physics~\cite{DeMille_2024} to controlled chemistry at absolute zero~\cite{Karman_2024,Yu_2022,Krems_2008}. A particularly compelling feature of these molecules is their large permanent electric dipole moment, which makes them promising candidates to advance quantum simulation of strongly correlated matter~\cite{Cornish_2024}. The study of tunable quantum gases, originally pioneered with atoms interacting via short-range forces~\cite{Bloch_2008, Chin_2010} and later expanded to dipolar magnetic atoms~\cite{Lahaye_2009, Chomaz_2022}, is now pushed into an entirely new regime by polar molecules. When polarized by external fields, they exhibit long-range, anisotropic dipole-dipole interactions that are substantially stronger than those of their magnetic atom counterparts. The ability to tune these interactions makes ultracold polar molecules an ideal platform for realizing extended Hubbard models \cite{Micheli_2006}, supersolid states of matter \cite{Schmidt_2022}, and engineering quantum computation architectures~\cite{DeMille2002}.

However, a persistent obstacle to realizing these applications has been the rapid, inelastic loss of molecules from optical traps, which inhibits the evaporative cooling required to reach quantum degeneracy. Even for chemically stable, non-reactive species, two-body loss rates approach the so-called universal limit, where any pair of molecules reaching the short range is lost with near-unity probability~\cite{Idziaszek2010, Ospelkaus_2010, Takekoshi_2014, Molony2014, Park2015, Guo_2016, Ye_2018, Bause_2023}. This phenomenon is attributed to the formation of long-lived two-body collision complexes~\cite{Mayle2012,Mayle2013}, which are subsequently depleted by background loss mechanisms such as photoexcitation by the trapping light~\cite{Christianen2019a,Bause_2023}. To overcome this challenge, various shielding techniques employing static electric and microwave fields~\cite{Avdeenkov_2006, Buchler2007, Gorshkov_2008, Cooper_2009, Karman2018, Lassabliere2018} have been developed. The underlying principle is to dress the molecules with external fields, engineering a long-range repulsive barrier that effectively prevents the molecules from reaching the short range. Shielding has since been realized experimentally in several molecular species~\cite{Valtolina_2020,Matsuda_2020,Li_2021,Anderegg_2021}, successfully suppressing two-body losses by orders of magnitude below the universal limit, allowing evaporative cooling and realization of degenerate molecular Fermi gases \cite{Valtolina_2020, Schindewolf_2022} and Bose-Einstein condensates \cite{Bigagli_2024, Shi_2026}.

Microwave shielding, one of the techniques for suppressing collisional loss, operates by dressing molecules with a circularly polarized ($\sigma^{+}$) microwave field blue-detuned from the lowest rotational transition (${j=0 \rightarrow 1}$). This field induces a rotating dipole moment in the molecules, which follows the rotation of the external field. As two molecules approach one another, the electric field generated by the approaching molecule eventually dominates over the external microwave field, and the dynamics are governed by resonant dipole-dipole interactions. When the molecules are prepared in the upper dressed state, this resonant interaction creates a strong repulsive barrier that prevents molecules from reaching the short range. {However, this approach introduces a critical trade-off. While stronger microwave fields improve two-body loss suppression, the time-averaged interaction between these rotating dipoles simultaneously creates an attractive potential well at long range.}
At high field strengths, this well becomes deep enough to support bound states, which triggers rapid three-body recombination and inherently limits the maximum shielding efficiency\cite{stevenson2024three,Yuan2025}. 

This limitation is overcome by introducing a second, linearly polarized ($\pi$) microwave field, {an approach conceptually analogous to a much earlier proposal using static field~\cite{Gorshkov_2008}}.
Because the $\pi$-field induces a dipolar interaction with the opposite sign to that of the $\sigma$-field, it effectively reduces the long-range attractive well and prevents the formation of field-linked bound states even at high microwave field intensities~\cite{Bigagli_2024, Karman2025, Deng_2025}. 
This allows applying stronger field strengths, thereby suppressing two-body loss rates to levels unattainable with a single field. Double microwave shielding led to the realization of Bose-Einstein condensates of polar molecules~\cite{Bigagli_2024, Shi_2026} and the observation of self-bound droplets in a strongly interacting gas of NaCs~\cite{Zhang_2026}. {The mechanisms of double microwave shielding were detailed theoretically in Ref.~\cite{Karman2025},
including a demonstration that loss suppression can be effective for a wide range of dipolar molecules.}
{This universality was recently formalized in Ref.~\cite{Dutta_2025} for the case of single-field microwave shielding.}

{While the observations of BEC and droplets prove the efficacy of double microwave shielding, these milestones were achieved without exploring the full parameter space available in double microwave shielding.
}
To fully exploit this technique, here, we comprehensively map the four-dimensional parameter space of double microwave shielding to identify the optimal operating regimes for future experiments. {Identifying these regimes is guided by several criteria.}
First, they must be strictly free of field-linked bound states to prevent three-body recombination. Second, they must {sufficiently} suppress the two-body loss rate, {ideally such} that the collisional lifetime of the gas exceeds its one-body lifetime, typically set by off-resonant scattering of the trapping light. {Finally, depending on the specific experimental goals, these regimes should either} maintain high elastic collision rates to enable efficient evaporative cooling, {or} offer broad tunability of the effective dipolar interactions {to facilitate quantum simulation}.

To systematically explore this four-dimensional parameter space, we frame our analysis using a set of dimensionless quantities. {Building on the recent universal description of single-field microwave shielding~\cite{Dutta_2025}, we express the scattering properties and microwave field parameters in terms of the characteristic dipolar length}
\begin{equation}
\label{eq:dipolar_length}
    a_{dd} =\frac{ \mu d^{2} } {4\pi \epsilon_{0} \hbar^{2}},
\end{equation}
and the corresponding dipolar energy scale
\begin{equation}
\label{eq:dipolar_energy}
    E_{dd} = \frac{\hbar^{2}}{2\mu a_{dd}^{2}}.
\end{equation} 
Here, $\mu$ denotes the reduced mass of the scattering system, and $d$ is the molecule's permanent dipole moment. {This approach is highly advantageous, as it effectively factors out the specific molecular mass and dipole moment. Consequently, a single set of dimensionless calculations applies universally to any chosen molecular species.} 

 We consider scattering in the low-temperature limit, where the two-body collisional loss rate coefficient $k$ for two identical bosons is related to the scattering length $a_{s}= \alpha - i\beta$ by
\begin{equation}
\label{eq:loss_rate} 
     k_{\mathrm{loss}} = \frac{2 d^2}{\epsilon_0 \hbar} \tilde{\beta}, 
\end{equation}
where $\tilde{\beta}=\beta/a_{dd}$ is the dimensionless imaginary part of the scattering length. {Crucially, $\tilde{\beta}$ is a universal function of the microwave field parameters expressed in units of the dipolar energy, entirely independent of the molecular species.} The physical prefactor in Eq.~\eqref{eq:loss_rate}, meanwhile, depends only on the molecule's dipole moment squared, $d^{2}$.

\begin{figure}
    \centering
    \includegraphics[width=\linewidth]{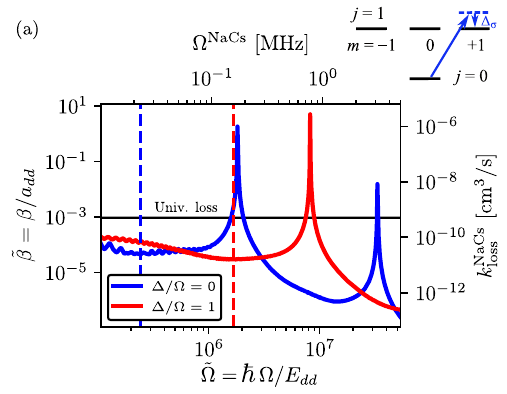}
    \includegraphics[width=\linewidth]{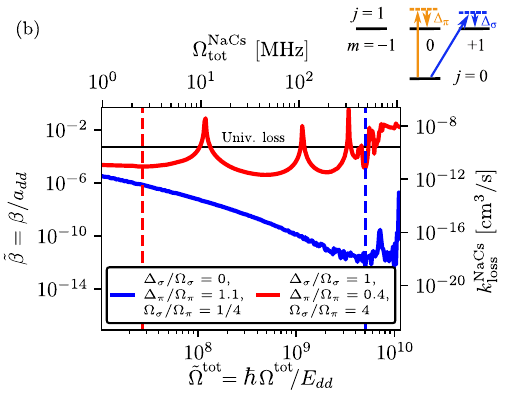}
    \caption{
    Two-body loss rates as a function of the applied microwave field strength for single ({a}) and double ({b}) microwave shielding. The universal dimensionless parameters, $\tilde{\Omega}$ and $\tilde{\beta}$, are shown on the bottom and left axes, while the corresponding physical quantities mapped for the NaCs molecule ($\Omega^{\mathrm{NaCs}}$ and $k^{\mathrm{NaCs}}_{\mathrm{loss}}$) are shown on the top and right axes. Solid lines represent the loss rates for selected configurations of the dimensionless field parameters. Vertical dashed lines in matching colors mark the corresponding optimal operating points ($\tilde{\Omega}_{\mathrm{opt}}$), which are located below the threshold of the first bound state, indicated by a sharp peak in the two-body loss rate. {Black horizontal lines indicate the universal loss rate for NaCs ($\sim 6.3 \times 10^{-10}\,\mathrm{cm}^3/\mathrm{s}$~\cite{Idziaszek2010}) corresponding to unshielded collisions.}}
    \label{fig:Omegascan}
\end{figure}

Figure~\ref{fig:Omegascan} illustrates the core principle behind our search for the optimal operating parameters. We first demonstrate this concept in the case of a single microwave dressing.
{For this scheme, the microwave field is parameterized by two frequencies: the detuning from the rotational transition $\Delta$ and the Rabi frequency $\Omega$. The relevant scaled} field parameters are the dimensionless Rabi frequency, $\tilde{\Omega} = {\hbar \Omega}/{E_{dd}}$, and the detuning-to-Rabi frequency ratio $\Delta / \Omega$. {This parametrization is physically intuitive: the ratio $\Delta/\Omega$ determines the composition of the field-dressed eigenstates, thereby fixing the induced dipole moment of shielded molecules. Conversely, $\tilde{\Omega}$ determines the overall energy scale.} Increasing $\tilde{\Omega}$ widens the energy gap between the dressed states and suppresses non-adiabatic transitions, which reduces the two-body loss rate.
However, this {overall scaling} simultaneously deepens the long-range attractive potential until it supports a field-linked bound state. Consequently, as illustrated in Fig.~\ref{fig:Omegascan}{(a)}, for any fixed ratio of detuning to Rabi frequency, $\Delta / \Omega$, there exists an optimal frequency, $\tilde{\Omega}_{\mathrm{opt}}$, located strictly below the threshold, $\tilde{\Omega}_{\mathrm{bound}}$, where the first field-linked bound state appears. This $\tilde{\Omega}_{\mathrm{opt}}$ minimizes the two-body loss rate for that specific field configuration. By systematically scanning across all possible detuning ratios and comparing these conditional minima, we {map this optimized loss rate, $\tilde\beta_{\mathrm{opt}}(\Delta/\Omega)$. Finding the minimum of this function identifies the} global optimum, yielding the absolute deepest loss suppression {achievable for} single-field microwave shielding.

Double microwave shielding introduces two additional degrees of freedom: the parameters of the $\pi$-polarized field ($\Omega_{\pi}, \Delta_{\pi}$). We parameterize the four-dimensional space using three dimensionless ratios: the relative field strength, $\Omega_{\sigma}/\Omega_{\pi}$, and the respective detuning ratios $\Delta_{\sigma}/\Omega_{\sigma}$ and $\Delta_{\pi}/\Omega_{\pi}$. The overall energy scale is set by the total effective Rabi frequency, $\Omega^{\mathrm{tot}} = \sqrt{\Omega_{\sigma}^{2}+\Omega_{\pi}^{2}}$, which we similarly scale as $\tilde{\Omega}^{\mathrm{tot}} = \hbar {\Omega}^{\mathrm{tot}} / E_{dd}$. 

To identify the optimal operating points within this four-dimensional parameter space, we apply the same underlying principle. For any given set of the three dimensionless parameters, we identify the bound state threshold, $\tilde{\Omega}^{\mathrm{tot}}_{\mathrm{bound}}$, and locate the corresponding optimal operating point, $\tilde{\Omega}^{\mathrm{tot}}_{\mathrm{opt}}$ below it (see Fig.~\ref{fig:Omegascan}{(b)}). {By mapping the resulting optimized loss rate, $\tilde{\beta}_{\mathrm{opt}}$, across the} entire four-dimensional {parameter space, we identify} the global minimum that provides the deepest two-body loss suppression.
Complementary to finding {this global minimum}, we explore the {broader} optimized parameter space---free of bound states and featuring sufficiently suppressed losses---to evaluate the prospects for efficient evaporative cooling and investigate the tunability of the effective dipolar interactions between shielded molecules.

Furthermore, by expressing our results in universal, dimensionless units, we can perform a comparative analysis across a wide range of molecular species. However, while the underlying dimensionless physics remains invariant, the scaling imposed by the respective dipolar lengths and energies drastically alters the physical field strengths required to reach these optimal regimes. To address this, we impose \textit{non-universal} constraints dictated by experimental feasibility. Specifically, we restrict our analysis to a realistic range of microwave field strengths: strong enough to ensure that microwave coupling dominates over {relevant energy scales}, such as tensor Stark shifts induced by optical traps {and molecular hyperfine structure}, yet low enough to be readily generated experimentally. 
By applying these practical constraints to {our universal parameter space scans, we address two critical questions for future experiments: which molecular species provide the most robust physical platforms, and how can one access the strongest effective dipolar interactions while successfully suppressing losses below critical thresholds.}

The manuscript is organized as follows. Section~\ref{sec:theory} outlines the theoretical framework behind the coupled-channel calculations of the scattering lengths, rate coefficients, and effective dipolar lengths. Section~\ref{sec:single_mwv} revisits the single-microwave shielding scheme to illustrate our search methodology and {translates the universal dimensionless results into physical units to evaluate experimental feasibility}. 
Section~\ref{sec:double_mwv} presents the comprehensive {analysis} of the double-microwave parameter space. 
In this section, we discuss the general {characteristics} of the optimal operating regimes (Sec.~\ref{subsec:overview}), evaluate the prospects for evaporative cooling with a special emphasis on the unique challenges posed by fermions (Sec.~\ref{subsec:evaporation}), and explore the tunability of the effective dipolar interactions (Sec.~\ref{subsec:tunability}). In Section~\ref{subsec:molecules}, we generalize our findings across a broad range of bialkali and coinage-metal molecules to identify the most promising candidates for future experiments. Finally, Section~\ref{sec:conclusions} provides a summary of our conclusions.

\section{Theoretical framework}
\label{sec:theory}
The theory underlying these calculations has been described in detail in Ref.~\cite{Karman2025}. Here we briefly outline the key points.

The molecules are modeled as rigid rotors with a permanent electric dipole moment, governed by the single-molecule Hamiltonian
\begin{equation}
    \label{eq:monomer_hamiltonian}
    \hat{H}^{(X)} = B_{\mathrm{rot}} \hat{j}^{2} + \hat{H}^{(X)}_{\mathrm{ac},\sigma} +\hat{H}^{(X)}_{\mathrm{ac},\pi}  .
\end{equation}
The first term describes the rotational kinetic energy of molecule $X$, where $B_{\mathrm{rot}}$ is the rotational constant and $\hat{j}$ is the angular momentum operator associated with the rotation of the molecular axis. The second and third terms describe the interactions with the external microwave fields
\begin{equation}
    \label{eq:mol-field_hamiltonian}
    \hat{H}^{(X)}_{\mathrm{ac},\nu} = -\frac{1}{2} \frac{E_{\nu}}{\sqrt{N_{0,\nu}}} \Bigl(\hat{d}_{\nu}^{(X)} \hat{a}_{\nu} + \hat{d}_{\nu}^{(X)\dagger} \hat{a}_{\nu}^{\dagger}\Bigr),
\end{equation}
where $\hat{a}_{\nu}^{\dagger}$ and $\hat{a}_{\nu}$ denote the raising and lowering operators for microwave field mode $\nu$. The fields themselves are described by the Hamiltonian
\begin{equation}
    \label{eq:field_hamiltonian}
    \hat{H}_{\mathrm{ac},\nu} = \hbar\omega_{\nu} \Bigl(\hat{a}_{\nu}^{\dagger}\hat{a}_{\nu}  - N_{0,\nu}\Bigr) .
\end{equation}
The two fields oscillate at different frequencies $\omega_{\nu}$ and possess different polarizations. For linear $\pi$ polarization, $\nu=0$, while for the circular $\sigma^{+}$ polarization, $\nu=+1$. The spherical components of the dipole moment operator, $\hat{d}_{\nu}$, are related to their Cartesian counterparts via $\hat{d}_{0} = \hat{d}_{z}$ and $\hat{d}_{\pm} = \mp (\hat{d}_{x}\pm i \hat{d}_{y})/\sqrt{2}$. The Rabi frequency for the $j=0 \to 1$ transition is given by $\hbar \Omega_{\nu} = d E_{\nu}/\sqrt{3}$, where $E_{\nu}$ is the corresponding electric field strength, $d$ is the permanent dipole moment of the molecule. The microwave frequencies are characterized by their detuning, $\Delta_{\nu} = \omega_{\nu} - \omega_{0,0;1,\nu}$ from the bare $j=0, m=0  \to 1, \nu $ transition. {Equation \eqref{eq:field_hamiltonian} includes a shift by $N_{0,\nu}$, which sets the zero of energy at a macroscopic reference photon number, a standard technique in the dressed-atom approach~\cite{cohenbook, Napolitano1997}.}

We omit hyperfine terms in the monomer Hamiltonian. Previous studies on shielded collisions of ultracold molecules~\cite{Karman2018, Karman2019, Bigagli_2023} concluded that nuclear spins act as spectator degrees of freedom at moderate magnetic fields, typically above 100~G.

The total Hamiltonian for the pair of colliding molecules in the center-of-mass frame and in the presence of two microwave fields is 
\begin{equation}
    \label{eq:dimer_hamiltonian}
    \hat{H} = -\frac{\hbar^{2}}{2\mu} \frac{d^{2}}{dR^{2}} + \frac{\hat{\ell}^{2}}{2\mu R^{2}} + \hat{H}^{(A)} + \hat{H}^{(B)} + \hat{H}_{\mathrm{ac},\sigma} + \hat{H}_{\mathrm{ac},\pi} + \hat{V},
\end{equation}
where $\mu$ is the reduced mass of the system and $R$ is the intermolecular distance. The first and second terms represent the radial and centrifugal relative kinetic energies, respectively. The subsequent terms describe the monomer and bare field Hamiltonians as defined in Eq.~\eqref{eq:monomer_hamiltonian} and~\eqref{eq:field_hamiltonian}. The final term represents the interaction between the two molecules, which we represent as the dipole-dipole interaction
\begin{equation}
\label{eq:dipole_interaction}
    \hat{V} = - \frac{d^{2}\sqrt{30}}{4\pi \epsilon_{0}R^{3}} \Bigl[ [{C}^{(1)}(\hat{r}^{(A)}) \otimes {C}^{(1)}(\hat{r}^{(B)})]^{(2)} \otimes {C}^{(2)}(\hat{R})\Bigr]^{(0)}_{0}.
\end{equation}
Here, $C^{(1)}(\hat{r}^{(X)})$ is the rank-1 tensor with spherical components given by Racah-normalized spherical harmonics depending on the polar coordinates of $\hat{r}^{(X)}$, $C^{(2)}(\hat{R})$ is an analogous rank-2 tensor depending on the polar angles of the intermolecular axis, $\hat{R}$, and 
\begin{equation}
    [{A}^{(k_{A})}  \otimes {B}^{(k_{B})}]^{(k)}_{q} = \sum_{q_{A}q_{B}} \hat{A}^{(k_{A})}_{q_{A}}
    \hat{B}^{(k_{B})}_{q_{B}} \langle k_{A} q_{A} k_{B} q_{B}| kq \rangle,
\end{equation}
is the $q$ spherical component of the rank-$k$ irreducible spherical tensor product of $\hat{A}$ and $\hat{B}$. These are tensors of rank $k_{A}$ and $k_{B}$, respectively, and $\langle k_{A} q_{A} k_{B} q_{B}| kq \rangle$ is a Clebsch-Gordan coefficient.

We use a fully uncoupled basis set. The monomer states are represented simply as eigenvectors of $\hat{j}^{2}$, denoted by $|j m\rangle$. The state of the microwave fields is described in the photon number basis, $|N_{\nu}\rangle$. Consequently, the basis functions describing a single molecule $X$ dressed by two microwave fields are $|j^{X} m^{X}\rangle |N_{\sigma} \rangle |N_{\pi}\rangle$. Matrix elements of the monomer Hamiltonian are straightforward to compute: the rotational term is strictly diagonal, and the matrix elements of the creation and annihilation operators are given by $\langle N | \hat{a}| N' \rangle = \delta_{N,N'+1} \sqrt{N}$, where $\delta$ is a Kronecker delta. In this representation,  the total number of photons in field $\nu$ is given by ${N_{\nu} + N_{0,\nu}}$, where $N_{0,\nu}$ denotes the reference number of photons. Note that for classical fields where $N_{0,\nu}$ is large, matrix elements of the Hamiltonians are independent of this reference number, {and the approach used here is equivalent to the Floquet description of a molecule interacting with a classical oscillating electric field~\cite{Guerin_1997}.}

Our calculations proceed by first constructing the monomer Hamiltonian, computing its eigenstates, and identifying the specific eigenstate in which the molecules are initially prepared. This corresponds to the upper field-dressed state, denoted as $|+\rangle$. This state is a linear combination of the bare states $|j=0, m=0\rangle |N_{\sigma} =0\rangle |N_{\pi} =0\rangle$, $|j=1, m=0\rangle |N_{\sigma} =0\rangle |N_{\pi} =-1\rangle$, and $|j=1, m=1\rangle |N_{\sigma} =-1\rangle |N_{\pi} =0\rangle$, where the precise amplitudes are determined by the specific detunings and Rabi frequencies of the applied microwave fields.

We then represent the dimer Hamiltonian in the product basis set that incorporates the basis vectors of both molecules, the partial wave basis set, $|\ell m_{\ell}\rangle$, describing the end-over-end rotation of the dimer, and the photon states:
\begin{equation}
    |j^{A} m^{A}\rangle |j^{B} m^{B}\rangle |\ell m_{\ell}\rangle |N_{\sigma} \rangle |N_{\pi}\rangle.
\end{equation}
After constructing the dimer Hamiltonian in this primitive basis, we adapt it to permutation symmetry of identical particles by applying the projection operator $1+\hat{P}$ (where $\hat{P}$ permutes molecules $A$ and $B$) {for identical bosons, or $1-\hat{P}$ for identical fermions,} and normalizing appropriately. We then determine the asymptotic basis set by numerically diagonalizing the Hamiltonian, excluding the molecule-molecule interaction term, for each independent $\ell, m_{\ell}$.
All matrices previously constructed in the parity-adapted primitive basis set, representing the monomer Hamiltonian, the centrifugal term, and the dipole-dipole interaction, are subsequently transformed into this permutation-adapted asymptotic representation, where all scattering calculations are performed. 

In our calculations, the molecular basis set typically includes only the rotational functions with $j=0$ and $j=1$. The photon basis set is limited to functions with $N_{\nu}$ between $-4$ and $2$ for both the $\sigma^{+}$ and $\pi$ field. The partial wave basis set includes functions {with $\ell$ up to 6}. Upon determining the asymptotic eigenbasis, we truncate it based on the asymptotic energy, retaining only functions within $\pm B_{\mathrm{rot}}/2$ of the initial state or excited by $4 B_{\mathrm{rot}}$. This limits the basis to only the set of nearly degenerate states in the vicinity of the initial state and the states that determine the rotational van der Waals interaction. The validity of this truncation is first tested and verified in Sec.~\ref{subsec:universality}, before being used in the rest of the work.  

The collisional loss and elastic rates are computed using coupled-channel scattering calculations as described in Refs.~\cite{Karman2018, Karman2019, Karman2020, Anderegg_2021, Schindewolf_2022, Chen_2023, Bigagli_2023, Bigagli_2024}. {To fully scan the parameter space, calculations are performed using the physical parameters of the NaCs molecule. The resulting observables and corresponding field parameters are subsequently scaled by the NaCs dipolar length and energy to yield dimensionless quantities.} We propagate two linearly independent sets of solutions of the coupled-channel equations using the renormalized Numerov method~\cite{Janssen2013} and impose capture boundary conditions at short range (${R_{\mathrm{min}} = 250\,a_{0}}$), which assume that all flux disappears at the inner boundary. At asymptotically large distances (${R_{\mathrm{max}} = 10^{5}\,a_{0}}$), we apply standard $S$-matrix boundary conditions, corresponding to the unit incoming flux in the entrance channel and the outgoing flux in all other open channels. We use an initial step size of $1\,a_{0}$, which is doubled several times at large intermolecular distances.

From the asymptotic $S$-matrix we extract the $s$-wave scattering length, $a_{s}$, as
\begin{equation}
    a_{s} = \lim_{E\to 0} \frac{1 - S_{i,\ell=0,m_{\ell}=0;i,\ell=0,m_{\ell}=0}(E)}{ik\Bigl(1 + S_{i,\ell=0,m_{\ell}=0;i,\ell=0,m_{\ell}=0}(E)\Bigr)}
\end{equation}
where $i$ is a shorthand notation for the initial state that the molecules are prepared in, $|+\rangle|+\rangle$. The scattering length {is} constant in the limit of low energies, which we confirm numerically by changing $E$ between 0.1~pK and {10~pK. All production calculations in this work are performed at a collision energy of $E=0.1\,\mathrm{pK}$ for NaCs.} The loss rate is computed from the imaginary part of the scattering length $a_{s} = \alpha - i \beta$, according to: 
\begin{equation}
\label{eq:loss_rate_again} 
     k_{\mathrm{loss}} = \frac{8 \pi \hbar }{\mu} {\beta} .
\end{equation}
{At finite collision energies, extracting the inelastic two-body loss rate directly from the imaginary part of the scattering length can overestimate the true loss~\cite{Mukherjee_2023}. This is because $\beta$ captures all flux leaving the incoming $s$-wave channel, which includes not only state-changing collisions, but also off-diagonal elastic scattering into higher partial waves (e.g., $\ell=0 \to 2$), driven by the dipole-dipole interaction. The magnitude of this elastic contribution scales as $\sim\sqrt{E} {a_{dd}^{\mathrm{eff}}}^{2}$, where $a^{\mathrm{eff}}_{dd}$ is the effective dipolar length for shielded molecules (see the next paragraph). This contribution is thus more prominent in single than in double microwave shielding, where effective dipolar interactions can be compensated. We confirm numerically that the true inelastic loss rate, extracted directly from the relevant $S$-matrix elements, is independent of temperature up $1\,\mathrm{nK}$ for single microwave shielding and double microwave shielding far from the compensation point, and agrees with the loss rate computed from $\beta$ at $0.1\,\mathrm{pK}$. Close to the compensation point, this equivalence holds up to at least $1\,\mu\mathrm{K}$ for NaCs.}

The corresponding rate for elastic collisions is given as the sum of the $s$-wave scattering length and the dipolar contribution~\cite{Bohn_2009}
\begin{equation}
\label{eq:elastic_rate} 
     k_{\mathrm{el}} =\langle v \rangle  \Bigl( 8 \pi a_{s}^{2} + \frac{32\pi}{45} \,{a^{\mathrm{eff}}_{dd}}^{2} \Bigr),
\end{equation}
where $\langle v \rangle = \sqrt{8k_{B}T/(\pi \mu)}$ is the mean relative collision energy at given temperature, and $a^{\mathrm{eff}}_{dd} = ({ \mu d^{2}_{\mathrm{eff}}})/({4\pi\epsilon_{0}\hbar^{2}})$ is the effective dipolar length for shielded molecules. 
In the case of double microwave field dressing, the effective induced dipole moment, $d_{\mathrm{eff}}$, depends on the detuning-to-Rabi frequency ratios ($\Delta_{\pi}/\Omega_{\pi}$, $\Delta_{\sigma} / \Omega_{\sigma}$) and the relative field strength ($\Omega_{\sigma} / \Omega_{\pi}$). We determine $d^{2}_{\mathrm{eff}}$ for each field configuration by extracting it from the effective interaction between the two dressed molecules~\cite{Karman2025}
\begin{equation}
\label{eq:induced_dipolar_interaction}
    V_{\mathrm{eff}}(\mathbf{R}) =- \frac{2 d_{\mathrm{eff}}^{2}}{4\pi\epsilon_{0}R^{3}} C_{2,0}(\hat{R}) ,
\end{equation}
where $C_{2,0}(\hat{R})$ quantifies the dependence of the interaction on the orientation of the intermolecular axis. Specifically, we isolate the effective dipole moment by taking the expectation value $V_{i,\ell=2,m_{\ell}=0;i,\ell=0,m_{\ell}=0}$ of the dipolar interaction [Eq.~\eqref{eq:dipole_interaction}] in the initial state the molecules are prepared in. Note that, since the expectation value vanishes within $\ell=0, m_{\ell} = 0$, we evaluate the matrix element between the $\ell = 0, m_{\ell} = 0$ and $\ell = 2, m_{\ell} = 0$ states to integrate out the angular dependence.

 \subsection{Universality}
\label{subsec:universality}
 \begin{figure*}
    \centering
    \includegraphics[width=0.89\linewidth]{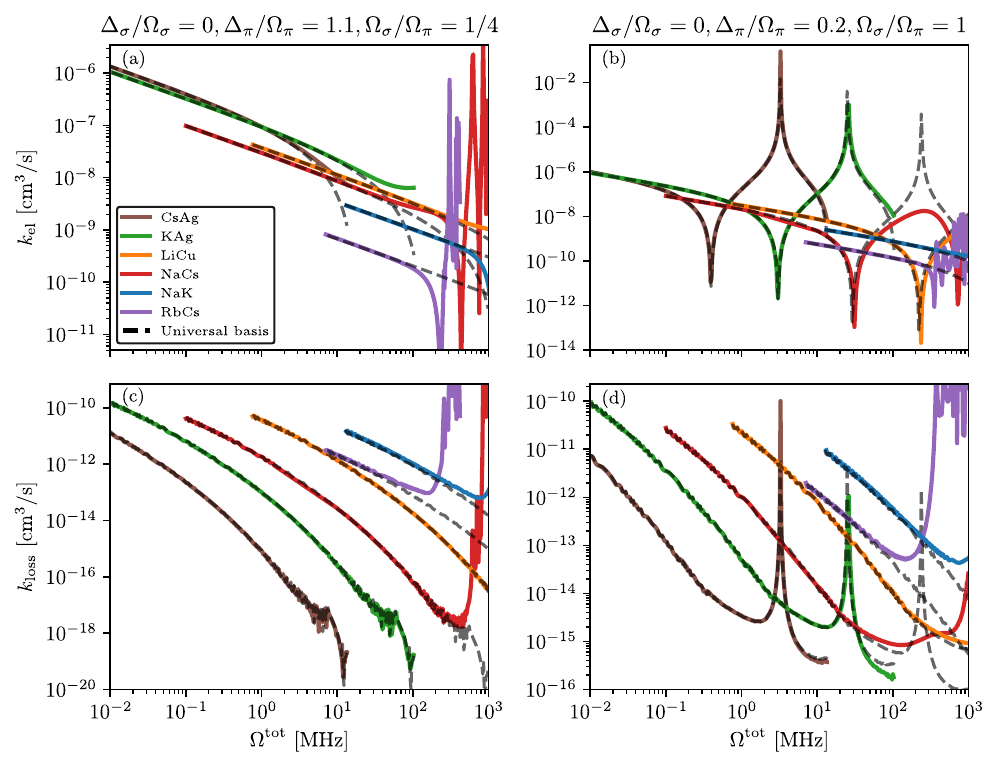}
    \caption{Limits of universality in double microwave shielding. Elastic (a, b) and loss (c, d) rate coefficients are shown for selected molecules for two selected double microwave shielding configurations. The universal baseline (dashed lines), obtained from NaCs calculations in the universal basis set (see text), is rescaled by the respective dipolar units for each species. Solid lines indicate coupled-channel calculations performed in an extended basis set including $j=2$ rotational states for each molecule.}
    \label{fig:universality_convergence}
\end{figure*}

Coupled-channel equations for shielded polar molecules take a universal form when expressed in dipolar units, {scaling lengths by $a_{dd} = { \mu d^{2} }/ (4\pi \epsilon_{0} \hbar^{2})$ and energies by $E_{dd}=\hbar^{2}/(2\mu a_{dd}^{2})$}~\cite{GonzalezMartinez2017, Dutta_2025}. 
In the low-field regime, where the Rabi frequencies and detunings are much smaller than the rotational constant, the collision dynamics are dominated by a small subset of near-degenerate {field-dressed} molecular states.
In this limit,  coupled-channel equations become essentially independent of specific molecular parameters. For single microwave shielding, Dutta \textit{et al.}~\cite{Dutta_2025} demonstrated that a minimal universal basis set restricted to rotational states $j=0, 1$---specifically comprising 10 pair states satisfying the resonance condition $j_{A} + j_{B} +  N_{\sigma}=0$---reproduces exact rate coefficients to within 5\% for field strengths ${\Omega}_{\sigma} / {B}_{\mathrm{rot}} \lesssim 0.02$. At stronger fields, {however}, the admixture of $j=2$ {rotational} states becomes significant. {Because the relative energy of these higher rotational states introduces an additional dimensionless parameter, $B_{\mathrm{rot}}/E_{dd}$, the scattering observables become species-dependent, and the universality is formally broken}.

Constructing an analogous universal basis set for double microwave shielding introduces a fundamental complication. While the single-field resonance condition restricts the relevant molecular states, the presence of two fields modifies this constraint to $j^{A}+j^B + N_{\sigma} + N_{\pi} =0$. Consequently, there is an infinite manifold of field-dressed states that satisfy this condition by exchanging photons between the two microwave fields, i.e. $(j^A,j^B,N_{\sigma},N_{\pi}) = (0,0,N,-N),\,(1,0,-N,N-1)$, and so forth. Although only a subset of these states directly couples to the initial state, this expanded manifold of states must be carefully truncated to capture the universal physics. In this work, by restricting the photon numbers $N_{\sigma}$ and $N_{\pi}$ to the range between $-4$ and $2$, limiting rotational states to $j=0,1$, and applying the resonance condition, we construct a minimal universal basis set comprising 87 dimer states that is entirely independent of any specific molecular constant. 

We perform all the extensive parameter space scans using the NaCs molecule. The resulting loss and elastic rate coefficients, effective dipole moments, and optimal field parameters are then mapped to other molecules by simple scaling via the respective dipolar length and energies. To rigorously justify this transferability, we establish the field limits within which the universality holds for double microwave shielding.

Figure~\ref{fig:universality_convergence} compares the elastic and loss rates computed using the universal basis set against exact, molecule-specific calculations for selected species with varying masses and permanent dipole moments. The universal baseline is shown alongside exact calculations performed independently for each molecule in an extended basis that explicitly includes the $j=2$ states and uses the species-specific molecular parameters listed in Tab.~\ref{tab:molecules}.

The universal scaling accurately describes both elastic and loss rate coefficients up to field strengths corresponding to a few percent of the respective molecule's rotational constant. While the exact breakdown point varies among species---ranging from roughly $2\%$ for KAg ($\sim 40$~MHz) to as high as $30\%$ for LiCu ($\sim 400$~MHz)---a conservative threshold of $\Omega/B_{\mathrm{rot}} \lesssim 0.02$ universally guarantees quantitative agreement. It is important to note that exceeding this threshold does not imply that shielding becomes ineffective, but rather that the dynamics become explicitly dependent on $B_{\mathrm{rot}}$ and the pure dipolar scaling breaks down. Therefore, whenever we transfer our NaCs-derived results to other species in subsequent sections, we strictly limit our analysis to physical Rabi frequencies that satisfy the conservative $\Omega/B_{\mathrm{rot}} \lesssim 0.02$ criterion.

{Conversely, at low Rabi frequencies, the microwave coupling becomes comparable to non-universal energy scales, such as molecular hyperfine structure and tensor Stark shifts from optical traps. Similar to the high-frequency limit, this does not imply that shielding becomes ineffective, but rather that the collision dynamics become influenced by these species-specific effects, causing the pure dipolar scaling to break down.}

\begin{table}
    \caption{Parameters for selected polar molecules studied in this work. Dipole moments ($d$) are given in Debye (D), and rotational constants ($B_{\mathrm{rot}}$) are given in~GHz.}
    \label{tab:molecules}
    \begin{tabular*}{\columnwidth}{@{\extracolsep{\fill}} l c c r }
        \hline\hline
        \textbf{Molecule}  & $d$ (D) & $B_{rot}$ (GHz) & \textbf{Ref.}   \\
        \hline
        $^{23}\mathrm{Na}^{40}\mathrm{K}$    & $2.72$  & $2.82$  &~\cite{Park2015}  \\
        $^{7}\mathrm{Li}^{63}\mathrm{Cu}$    & $5.05$  & $15.76$ &~\cite{Smialkowski2021} \\
        $^{39}\mathrm{K}^{107}\mathrm{Ag}$   & $8.5$   & $2.0$   &~\cite{Smialkowski2021}  \\
        $^{23}\mathrm{Na}^{133}\mathrm{Cs}$  & $4.6$  & $1.74$  &~\cite{Dagdigian_1972} \\
        $^{87}\mathrm{Rb}^{133}\mathrm{Cs}$  & $1.225$ & $0.49$  &~\cite{Molony2014} \\
        $^{133}\mathrm{Cs}^{107}\mathrm{Ag}$ & $9.75$  & $0.806$ &~\cite{Smialkowski2021}  \\
        \hline\hline
    \end{tabular*}
\end{table}

\subsection{Microwave field ellipticity}
\label{subsec:ellipticity}
The polarization of the microwave fields enters the Hamiltonian through the spherical dipole components, $\hat{d}_{\nu}$, in Eq.~\eqref{eq:mol-field_hamiltonian}. For perfectly linear $\pi$ and circular $\sigma^{+}$ polarizations, $\nu=0$ and $\nu=+1$, respectively. In practice, however, achieving perfectly circular and linear polarization of the microwave fields at the frequencies and intensities needed for shielding is challenging. To account for this, the generalized dipole components interacting with the $\sigma$ and $\pi$ fields can be written as
\begin{align}
    \begin{split}
    \hat{d}_{\sigma} &= \hat{d}_{+1} \cos\xi - \hat{d}_{-1}\sin\xi,   \\
    \hat{d}_{\pi} &= \hat{d}_{0}\cos\chi + \hat{d}_{+1} \sin\chi\cos\theta  - \hat{d}_{-1}\sin\chi \sin\theta.
    \end{split}
\end{align}
For small angles $\xi$ and $\chi$, the fields closely approximate perfectly circular and linear polarizations. Experimentally, polarization purities corresponding to ellipticities of a few degrees are routinely achievable \cite{Schindewolf_2022,Bigagli_2023,Bigagli_2024,Chen_2023,Lin2023}. For the $\pi$ field, the additional phase angle $\theta$ determines the nature of the transverse components, quantifying whether the total $\pi$ field polarization is purely elliptical, perfectly linear but tilted away from the quantization $z$-axis, or a combination of both.

Imperfect polarization affects the spatial anisotropy of the induced dipole-dipole interactions. If the angular dependence of the interactions induced by the $\sigma$ and $\pi$ fields is not exactly identical, the dipolar interaction cannot be fully compensated simply by changing the detunings and relative Rabi frequencies~\cite{Karman2025}.
Previous studies showed that, {compared to the perfectly circular and linear polarization case}, experimentally realizable ellipticities can increase losses by over an order of magnitude in single microwave shielding~\cite{Karman2019,Karman2025}, but predicted a much weaker dependence on polarization in the double microwave shielding case.

To quantify the robustness of our double microwave shielding parameter scan against these imperfect polarizations, we compute the loss rates for selected field configurations assuming realistic ellipticities~\cite{Zhang_2026}. Specifically, we set the ellipticity of the $\sigma^{+}$ field to $\xi=0.5^{\circ}$. For the $\pi$ field, we introduce a $\sigma^{+}$ contamination component of $\sin(4^{\circ})$ and a $\sigma^{-}$ component of $\sin(1^{\circ})$ by setting $\sin\chi\cos\theta = \sin(4^{\circ})$ and $\sin\chi\sin\theta = \sin(1^{\circ})$.

\begin{figure}
    \centering
    \includegraphics[width=\linewidth]{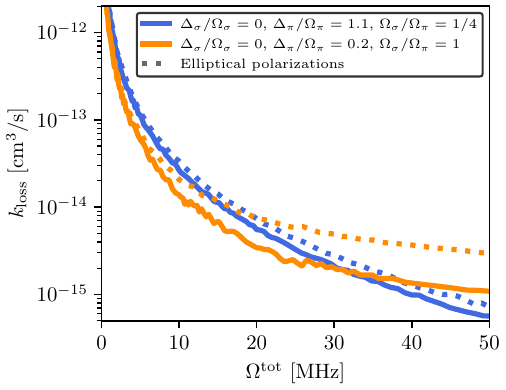}
    \caption{Influence of elliptical microwave polarizations on computed loss rates as a function of the total Rabi frequency, $\Omega^{\mathrm{tot}}$, for two selected field configurations. The inclusion of experimentally realistic ellipticities (see text for details) increases the baseline loss but preserves the qualitative shielding behavior and the location of optimal operating regimes. Results are computed for NaCs molecules.}
    \label{fig:ellipticity}
\end{figure}
Figure~\ref{fig:ellipticity} presents the influence of these imperfect polarizations on the NaCs loss rates for the two field configurations shown previously in Fig.~\ref{fig:universality_convergence}. As in Fig.~\ref{fig:universality_convergence}, we focus on the scaling of the loss rate with the total Rabi frequency, $\Omega^{\mathrm{tot}}$, which is the central optimization parameter in this work.

Crucially, the qualitative dependence of the loss rate on $\Omega^{\mathrm{tot}}$ remains unchanged: in the absence of field-linked bound states, the loss rate monotonically decreases with increasing field strength. The imperfect polarization introduces a relatively minor penalty, {generally increasing losses by a factor of 2 to 5, with this effect becoming slightly more pronounced as the field strength approaches the threshold for bound-state formation.} 

\section{Single microwave shielding}
\label{sec:single_mwv}
To illustrate our methodology for mapping the parameter space, we first address the single-microwave shielding. This configuration is characterized by a Rabi frequency $\Omega$ and the detuning $\Delta$ of a $\sigma^{+}$-polarized field {from the $j=0\to 1$ rotational transition. To ensure the results are universally applicable across different molecules, we express these parameters in dimensionless units, scaled by the characteristic dipolar energy, $E_{dd}$ [Eq.~\eqref{eq:dipolar_energy}]. The parameter space is thus described by the detuning-to-Rabi frequency ratio, $\Delta/\Omega$, which determines the composition of the field-dressed eigenstates, and hence the induced dipole moment of shielded molecules, and dimensionless Rabi frequency, $\tilde{\Omega} = \hbar \Omega / E_{dd}$, which sets the overall energy scale.}

{Our systematic scan of the parameter space, as illustrated previously in Fig.~\ref{fig:Omegascan}(a), proceeds as follows: for any fixed detuning-to-Rabi frequency ratio, we locate the optimal dimensionless Rabi frequency, $\tilde{\Omega}_{\mathrm{opt}}$, which minimizes the two-body loss rate, $\tilde{\beta}$ [Eq.~\eqref{eq:loss_rate}], while remaining below the threshold $\tilde{\Omega}_{\mathrm{bound}}$, where the first field-linked bound state appears. By repeating this procedure across different detuning ratios, we map the function $\tilde{\beta}_{\mathrm{opt}}(\Delta/\Omega)$. Finding the minimum of this function identifies the global minimum of the two-body loss rate achievable in the absence of field-linked bound states.}

\begin{figure}
    \centering
    \includegraphics[width=\linewidth]{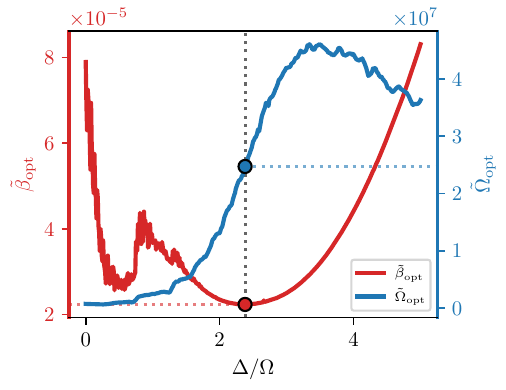}
    \caption{The lowest loss rate, expressed in terms of the dimensionless imaginary scattering length, $\tilde{\beta}_{\mathrm{opt}} = \beta_{\mathrm{opt}}/a_{dd}$, achievable for a fixed ratio of microwave parameters, $\Delta /\Omega $, under the constraint that no field-linked bound states are supported (red line, {left} vertical axis). The corresponding optimal Rabi frequency that minimizes the loss for each dimensionless configuration, $\tilde{\Omega}_{\mathrm{opt}} = \hbar \Omega / E_{dd}$, is presented as a blue line with values shown on the {right} vertical axis. The circular markers and vertical dotted line indicate the global optimal operating point at $\Delta/\Omega = 2.38$.}
    \label{fig:single_mwv_best}
\end{figure}

Figure~\ref{fig:single_mwv_best} presents this optimized dimensionless loss rate, $\tilde{\beta}_{\mathrm{opt}}$, and the corresponding optimal Rabi frequency, $\tilde{\Omega}_{\mathrm{opt}}$, as a function of $\Delta /\Omega$. It establishes the existence of a {unique, universally optimal} operating point for $\Delta/\Omega = 2.38$, $\tilde{\Omega}_{\mathrm{global}} = 2.47 \times 10^{7}$ corresponding to $\tilde{\beta}_{\mathrm{global}} = 2.23 \times 10^{-5}$.

\begin{figure*}
    \centering
    \includegraphics[width=0.49\linewidth]{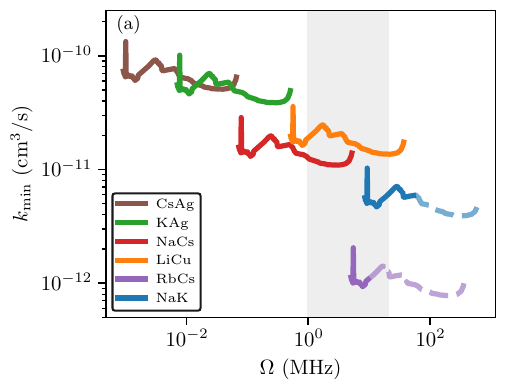}
    \includegraphics[width=0.49\linewidth]{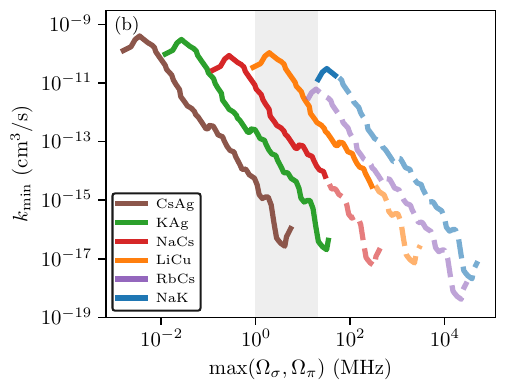}
    \caption{Minimum loss rate achievable {at any fixed} Rabi frequency in the single {(a)} and double {(b)} microwave shielding case. Highlighted is the region of experimental interest, i.e., $\Omega$ ranging between 1 and 20~MHz. The curves transition from solid to dashed where the Rabi frequency exceeds 2\% of the respective molecule's rotational constant. In this regime, state mixing with rotational levels $j > 1$ becomes non-negligible, and  {scaling results obtained with the universal basis set} is not expected to hold.
    Consequently, the dashed lines should be interpreted as qualitative estimates rather than exact quantitative predictions.}
    \label{fig:loss_binning}
\end{figure*}

To express this universal result practically, we {translate the optimized loss curve, $\tilde{\beta}_{\mathrm{opt}}(\Delta/{\Omega})$, into SI} units. 
{Specifically, Fig.~\ref{fig:loss_binning}(a) presents the absolute lowest two-body loss rate, $k_{\mathrm{min}}$, achievable at any fixed physical Rabi frequency, $\Omega$. Because the conversion factors depend on molecular mass and dipole moment, this translation separates the single universal curve into distinct, shifted, molecule-specific results. The vertical shift is governed by the $d^{2}$ scaling of the physical loss rate, $k_{\mathrm{loss}} \propto d^{2} \tilde{\beta}$ [Eq.~\eqref{eq:loss_rate}]. The horizontal shift reflects the scaling of the Rabi frequency, $\Omega = \tilde{\Omega} E_{dd}/\hbar \propto \tilde{\Omega} \mu^{-3} d^{-4}$.}

{This extreme $d^{-4}$ scaling determines which part of the $k_{\mathrm{min}}(\Omega)$ curve falls within the bounds of experimental feasibility. In Fig.~\ref{fig:loss_binning}(a), we highlight the $1-20$~MHz range}: fields of this magnitude are large enough to ensure that the microwave dressing dominates over tensor Stark shifts induced by optical dipole traps {and molecular hyperfine structure}, yet low enough to be readily implemented experimentally (we note that this is a conservative bound, as fields with higher intensities have been developed for shielding~\cite{Bigagli_2023}). Interestingly, we find that weakly dipolar molecules, such as RbCs ($d = 1.2\,\mathrm{D}$), achieve significantly lower absolute loss rates at experimentally feasible field range, compared to strongly dipolar species like NaCs ($d = 4.6\,\mathrm{D}$). For example, at $\Omega = 5$~MHz, the lowest achievable loss rate is an order of magnitude lower than that for NaCs. This remains true despite RbCs operating at Rabi frequencies far below the theoretical optimum (${k_{\mathrm{min}}= 8\times 10^{-13}}$~cm$^{3}$/s at $\Omega = 192$~MHz).

This finding {refines} recent interpretations of universality, which {initially suggested} that heavy, strongly dipolar molecules are favored for microwave shielding~\cite{Dutta_2025}. That previous conclusion arises from focusing solely on the scaling of the dimensionless Rabi frequency, $\tilde{\Omega} \propto \Omega  \mu^3 d^4$. It is indeed true that to achieve a \textit{fixed degree} of dimensionless shielding $\tilde{\Omega}$, a weakly dipolar species like RbCs requires an electric field roughly $10^2$ times larger than a strongly dipolar one like NaCs. However, {evaluating actual experimental feasibility requires accounting for two additional factors. First,} the absolute loss rate $k_{\mathrm{loss}}$ is the product of the dimensionless $\tilde{\beta}$ and a prefactor scaling as $d^{2}$. This means that while a much stronger field is indeed required to reach the same dimensionless $\tilde{\beta}$, the physical loss $k_{\mathrm{loss}}$ in SI units is suppressed by the $d^{2}$ factor. {Second, the formation of field-linked bound states imposes an upper limit on the dimensionless Rabi frequency, $\tilde{\Omega}$. For strongly dipolar molecules, this limit is reached at very low physical Rabi frequencies, drastically restricting the usable range of field strengths. Conversely, while fields with Rabi frequencies in the tens of MHz are not high enough for weakly dipolar molecules to reach their absolute global optimum, they still allow these molecules to suppress losses to the level that is only slightly higher than the theoretical minimum.}

\section{Double microwave shielding}
\label{sec:double_mwv}

\subsection{{Overview} of the parameter space}
\label{subsec:overview}

We now move on to {the double microwave shielding scheme~\cite{Karman2025}, which employs both a circularly ($\sigma^{+}$) and a linearly ($\pi$) polarized microwave fields. This configuration is parametrized by two Rabi frequencies ($\Omega_{\sigma}, \Omega_{\pi}$) and two detunings ($\Delta_{\sigma}, \Delta_{\pi}$) of the respective fields.} 

{To systematically explore this four-dimensional parameter space, we use an approach similar to the one used in the single-field case. First, we note that the composition of the field-dressed eigenstates, and the induced dipole moment of shielded molecules is determined by three dimensionless ratios: the relative field strength, $\Omega_{\sigma}/\Omega_{\pi}$, and the respective detuning ratios, $\Delta_{\sigma}/\Omega_{\sigma}$ and $\Delta_{\pi}/\Omega_{\pi}$. The overall energy scale is governed by the effective total Rabi frequency, $\Omega^{\mathrm{tot}} = \sqrt{\Omega_{\sigma}^{2}+\Omega_{\pi}^{2}}$. To ensure universality, we scale this total Rabi frequency by the dipolar energy, yielding the dimensionless $\tilde{\Omega}^{\mathrm{tot}} = \hbar {\Omega}^{\mathrm{tot}} / E_{dd}$.}

{With the parameter space defined, we perform a systematic scan analogous to the single-field case. For any fixed combination of the three dimensionless ratios, $\Omega_{\sigma}/\Omega_{\pi}$, $\Delta_{\sigma}/\Omega_{\sigma}$ and $\Delta_{\pi}/\Omega_{\pi}$, we scan the dimensionless total Rabi frequency, to locate its optimal value, $\tilde{\Omega}^{\mathrm{tot}}_{\mathrm{opt}}$. This optimum minimizes the two-body loss rate while remaining below the threshold $\tilde{\Omega}^{\mathrm{tot}}_{\mathrm{bound}}$, where the first field-linked bound state appears. By evaluating the optimized two-body loss rate, $\tilde{\beta}_{\mathrm{opt}}$, across a broad range of combinations of these three ratios, we construct a comprehensive {overview} of the parameter space in the double microwave shielding case. Throughout these calculations, we assume ideal polarizations of the $\sigma^{+}$ and $\pi$ fields.}

\begin{figure*}
    \centering
    \includegraphics[width=0.9\linewidth]{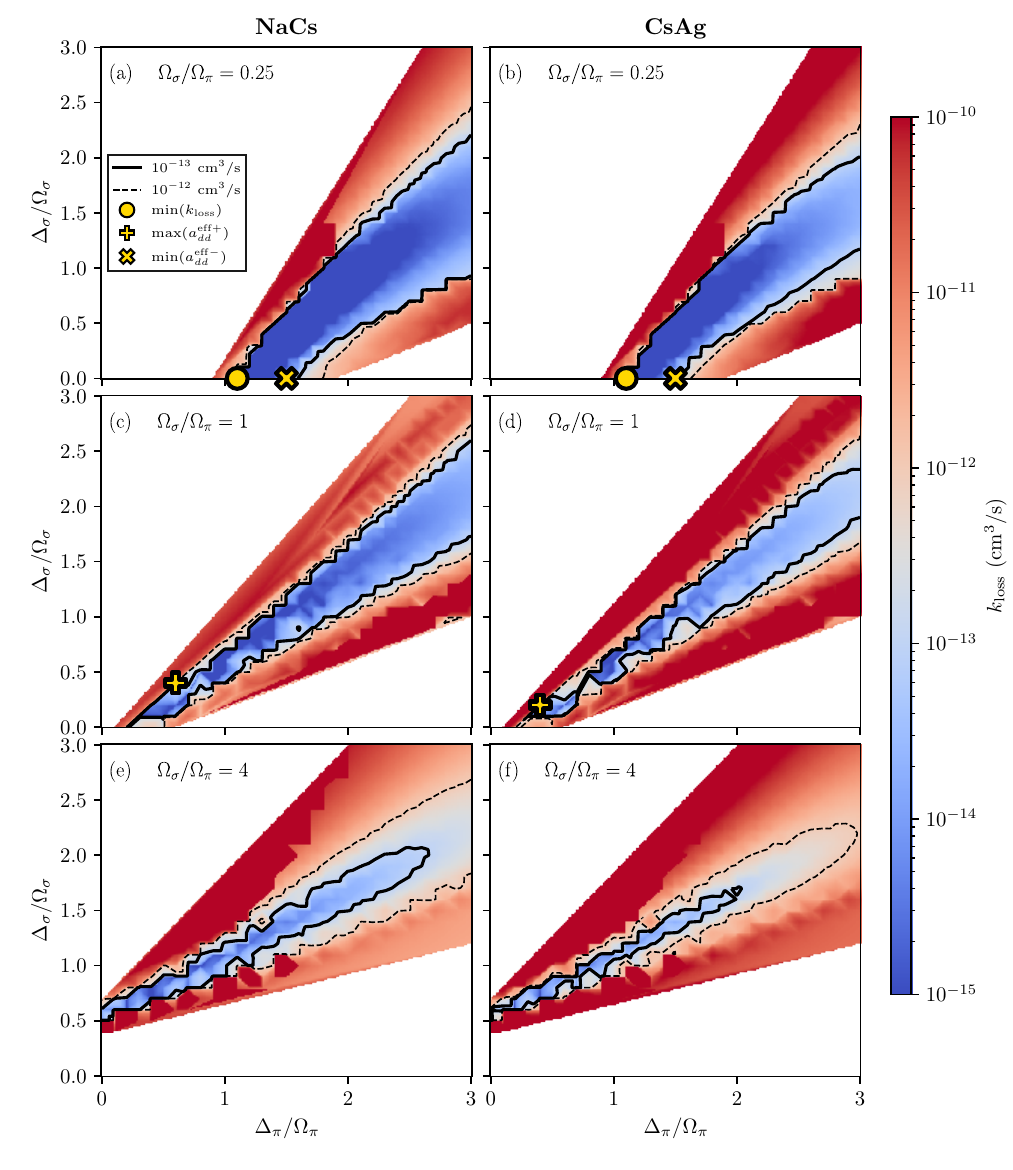}
    \caption{Example 2D cuts of the parameter space for NaCs (a, c, e) and {CsAg} (b, d, f) at selected Rabi frequency ratios ($\Omega_{\sigma}/\Omega_{\pi}$). The color map indicates the two-body loss rate $k_{\mathrm{loss}}$ (in~cm$^{3}$/s), with darker regions corresponding to stronger suppression. The solid black contour encloses a zone where losses are below the threshold of $10^{-13}$~cm$^{3}$/s. The dashed contour marks $10^{-12}$~cm$^{3}$/s. White regions indicate parameter space excluded by the appearance of bound states which limit shielding efficiency. Gold markers indicate the maximum effective dipolar lengths ($a^{\mathrm{eff}}_{dd}$) achievable within the constraint that $k<10^{-13}$~cm$^{3}$/s. The plus ($+$) and cross ($\times$) indicate the maximum positive and negative dipolar lengths, respectively, while the circle $\circ$ shows the configuration that minimizes the two-body loss rate. See text for details.}
    \label{fig:loss_dipole}
\end{figure*}

Figure~\ref{fig:loss_dipole} presents 2D cuts of this parameter space for NaCs (left column) and {CsAg} (right column). Each row corresponds to a fixed ratio of Rabi frequencies, representing a $\pi$-dominated ($\Omega_{\sigma} / \Omega_{\pi}=1/4$), balanced ($\Omega_{\sigma} / \Omega_{\pi}=1$), and $\sigma$-dominated ($\Omega_{\sigma} / \Omega_{\pi}=4$) field combinations. The heat maps display the minimum two-body loss rate achievable at the optimal total Rabi frequency, $\tilde{\Omega}^{\mathrm{tot}}_{\mathrm{opt}}$, as a function of the dimensionless detunings ($\Delta_{\sigma} / \Omega_{\sigma}$ and $\Delta_{\pi} / \Omega_{\pi}$). {In Fig.~\ref{fig:loss_dipole}, we do not yet impose any experimental upper bound on the field strength, allowing $\tilde{\Omega}^{\mathrm{tot}}_{\mathrm{opt}}$ to reach the theoretical optimal value.} Contour lines marking regions where losses are suppressed below $10^{-12}$~cm$^{3}$/s (dashed) and $10^{-13}$~cm$^{3}$/s (solid). {These two limits indicate the level of loss suppression sufficient for the two-body lifetime to exceed typical one-body lifetimes in ultracold experiments. {Assuming typical molecular gas densities of $n\sim 10^{12}\,\mathrm{cm}^{-3}$~\cite{Bigagli_2024,Yuan2025,Shi_2026}, suppressing the loss rate coefficient to $10^{-13}$~cm$^{3}$/s ensures collisional lifetimes on the order of 10 seconds, safely exceeding one-body lifetimes.}} Markers indicate the configurations that maximize the effective dipolar length to its largest positive ($+$) or negative ($\times$) value for each molecule. {While we place no upper limit on the fields, we do enforce a practical lower bound:} the white regions represent parts of the parameter space where the first field-linked bound state appears at exceedingly low Rabi frequencies ($\Omega < 100$~kHz), which are comparable to the tensor Stark shifts of typical optical traps.

The most prominent feature of these {2D cuts} is the characteristic shape of the region that effectively suppresses the three-body loss, resembling a cone that broadens at larger detunings. The center of the cone tracks the vicinity of the compensation point, where the $\pi$-field effectively {reduces the depth of} the long-range attractive well created by the $\sigma$-field. As the balance between the field strength changes, this compensation condition shifts systematically. For instance, if the fields are $\pi$-dominated (top panels in Fig.~\ref{fig:loss_dipole}), the $\pi$-field must be detuned to compensate the weaker $\sigma$-field, shifting the optimal shielding region to $\Delta_{\pi} / \Omega_{\pi} = 1-2$. Conversely, for $\sigma$-dominated fields (bottom panels in Fig.~\ref{fig:loss_dipole}), the region moves towards $\Delta_{\sigma} / \Omega_{\sigma} > 0.5$.

Because this compensation {reduces the depth of} the long-range attractive potential, the appearance of the first bound state is pushed to extremely high Rabi frequencies. This effectively removes the primary constraint on the field strength, allowing $\tilde{\Omega}^{\mathrm{tot}}$ to be increased by orders of magnitude. Indeed, within the $\pi$-dominated field configuration ($\Omega_{\sigma} / \Omega_{\pi} = 1/4, \Delta_{\sigma}/\Omega_{\sigma} = 0, \Delta_{\pi}/\Omega_{\pi} = 1.1$) we find a theoretical global minimum for two-body loss, with the optimal dimensionless Rabi frequency of ${\tilde{\Omega}^{\mathrm{tot}}_{\mathrm{global}} = 1.7 \times 10^{10}}$ . At this specific combination of parameters, the dimensionless loss rate is $\tilde{\beta}_{\mathrm{global}} = 2.8 \times 10^{-13}$, eight orders of magnitude smaller than the absolute minimum loss achievable in the single microwave field.

There is, however, a fundamental limit to exploiting this theoretical minimum in practice. Once the dimensionless optimum ($\tilde{\Omega}^{\mathrm{tot}} \sim 10^{10}$) is mapped back to SI units, the required physical Rabi frequencies for many species are on the order of gigahertz. This presents a twofold problem. First, microwave fields of such magnitude are experimentally unfeasible to generate. Second, fields of this strength break the universality assumption, which relies on restricting the Hilbert space to the ground and first excited rotational states ($j=0,1$, see Sec.~\ref{subsec:universality}). This approximation remains valid only when the Rabi frequency is much smaller than the rotational constant (see Fig.~\ref{fig:universality_convergence}). Because the rotational constants of bialkali molecules also lie in the~GHz range, pushing $\Omega$ to the theoretical global minimum violates this assumption for most species, as mixing with higher rotational states will inevitably occur.

The translation from the dimensionless framework to physical units is illustrated in Fig.~\ref{fig:loss_binning}(b). For selected alkali and coinage-metal dimers, it shows the minimum loss rate achievable at a fixed limiting Rabi frequency, $\max(\Omega_{\sigma}, \Omega_{\pi})$. The absolute minimum of the loss rate scales as $d^2$, which initially suggests (as in the single-field case) that weakly dipolar molecules might be superior. Indeed, the minimum loss for RbCs is $\sim 6 \times 10^{-20}$~cm$^3$/s, two orders of magnitude lower than that of CsAg. However, achieving this optimum for RbCs requires $\Omega_{\sigma} \approx 9.9$~GHz and $\Omega_{\pi} \approx 40$~GHz, which is far beyond experimental feasibility and exceeds the rotational constant ($B_{\mathrm{rot}} \approx 0.5$~GHz). The breakdown of universality is explicitly indicated in Fig.~\ref{fig:loss_binning} by the transition from solid to dashed lines when $\Omega/B_{\mathrm{rot}}> 0.02$.

\begin{figure*}
    \centering
    \includegraphics[width=0.8\linewidth]{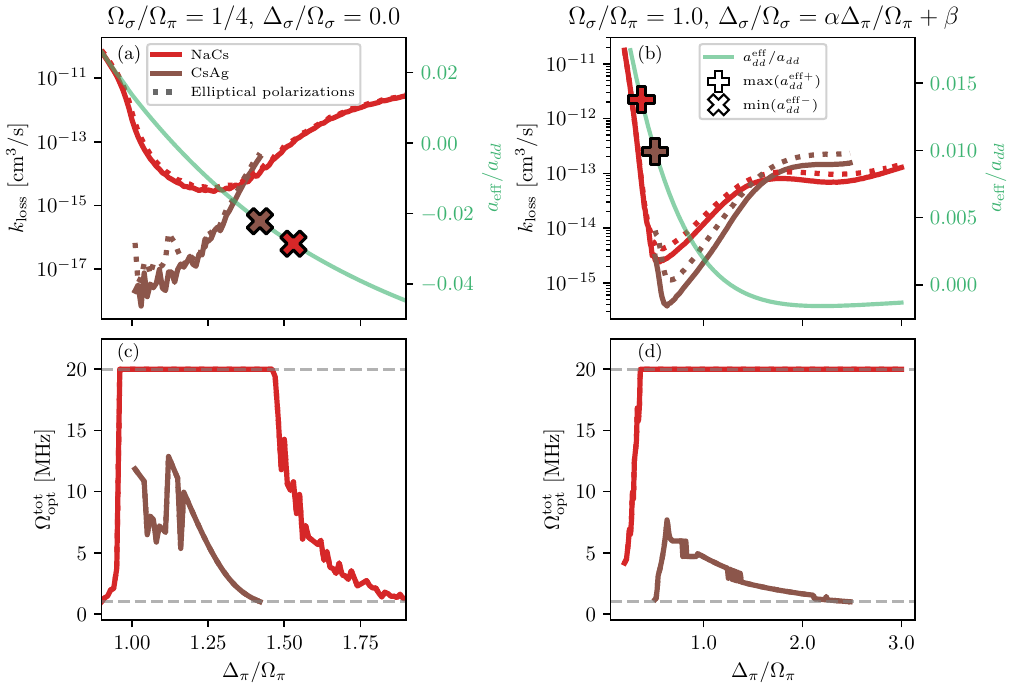}
    \caption{ 
    Example 1D cuts of the parameter space showed in Figure~\ref{fig:loss_dipole}, detailing two-body loss rates and effective dipolar lengths. The left column ({panels a, c}) present a horizontal cut through the $\pi$-dominated field configuration ($\Omega_{\sigma}/\Omega_{\pi} = 1/4$) at $\Delta_{\sigma}/\Omega_{\sigma} = 0$. The right column (panels b, d) follows a diagonal cut through the balanced field strengths ($\Omega_{\sigma}/\Omega_{\pi} = 1$) along $\Delta_{\sigma}/\Omega_{\sigma} = \alpha \Delta_{\pi}/\Omega_{\pi} + \beta$, where $\alpha = 0.68, \,\beta = -0.072$. In the top panels (a, b), solid curves (left axis) represent the minimum loss rates for NaCs and CsAg achievable with Rabi frequencies limited to $1-20$~MHz. Dashed curves show the corresponding loss rates when realistic microwave field ellipticity is included. The solid green line (right axis) tracks the effective dipolar interaction, scaled by the universal dipolar length ($a_{\mathrm{eff}}/a_{dd}$). {Markers, color-coded to match the respective molecular species,} highlight the limits of dipolar tunability achievable within a collisionally stable gas: $\times$ and $+$ indicate the maximum negative and positive effective dipolar lengths, respectively, that can be reached while maintaining a loss rate strictly below $10^{-13}$~cm$^3$/s. The bottom panels (c, d) display the corresponding overall Rabi frequencies ($\Omega^{\mathrm{tot}} = \sqrt{\Omega_{\sigma}^{2} + \Omega_{\pi}^{2}}$) applied at each point, with dashed gray lines marking the 1 and 20~MHz limits.}
    \label{fig:1d_cuts}
\end{figure*}

Conversely, for heavy, strongly dipolar species like {CsAg}, the scaling factor of $\tilde{\Omega} \propto \Omega \mu^3 d^4$ shifts the optimal operating point into the feasible, universal regime. For CsAg, the universal optimum corresponds to {$\Omega_{\sigma} \approx 2.7$~MHz and $\Omega_{\pi} \approx 10.6$~MHz. While the resulting loss rate ($\sim 7 \times 10^{-19}$~cm$^3$/s)} is higher than the theoretical limit for RbCs, it is physically realizable, fulfills the $\Omega/B_{\mathrm{rot}} \lesssim 0.02 $ condition, and remains orders of magnitude below the {$10^{-13}$~cm$^3$/s threshold established above.}
Thus, we arrive at the opposite conclusion to the single-field case: when restricted to experimentally feasible Rabi frequencies ($\Omega \sim 10$~MHz), heavy, strongly dipolar molecules offer the most robust path to extreme loss suppression in the double-microwave scheme.

While Figure~\ref{fig:loss_dipole} provides a {broad overview} of the optimal shielding regimes, {one might naturally ask how this translates to practical operation when field strengths are capped. Rather than presenting separate capped 2D cuts, we note that the color scale in Fig.~\ref{fig:loss_dipole} is purposefully saturated at $10^{-15}$ cm$^{3}$/s, which corresponds to the deepest loss suppression achievable for NaCs under a realistic 20~MHz limit on Rabi frequencies. Consequently, the darkest regions in Fig.~\ref{fig:loss_dipole} already show the most promising parts of the parameter space under realistic experimental constraints.} 

To more quantitatively examine these {regimes}, Figure~\ref{fig:1d_cuts} presents two selected 1D cuts through the parameter space. Let us reiterate that each point along these cuts corresponds to a total Rabi frequency, $\Omega^{\mathrm{tot}}_{\mathrm{opt}}$, specifically optimized to minimize the two-body loss rate for that exact set of dimensionless parameters. {However,} following the discussion of single microwave shielding, we {now explicitly} restrict $\Omega^{\mathrm{tot}}_{\mathrm{opt}}$ to field strengths within $1-20$~MHz.

The left column of Figure~\ref{fig:1d_cuts} [panels (a) and (c)] presents a horizontal cut through the $\pi$-dominated field configuration [$\Omega_{\sigma} / \Omega_{\pi} = 1/4$, see Fig.~\ref{fig:loss_dipole} (a, b)] at $\Delta_{\sigma} / \Omega_{\sigma} = 0$. This specific slice passes through the vicinity of the theoretical global minimum for loss suppression ($\Delta_{\pi} / \Omega_{\pi} = 1.1$). Here, we can explicitly see the impact of imposing non-universal constraints (i.e., constraints in real units) on the field strength. The physical origin of the resulting loss profiles is revealed in panel (c), which tracks the optimized overall Rabi frequency, $\Omega^{\mathrm{tot}}_{\mathrm{opt}}$. Because reaching the theoretical global minimum requires very high field strengths, the optimal frequency for NaCs exceeds the upper limit, saturating $\Omega^{\mathrm{tot}}_{\mathrm{opt}}$ at 20~MHz between $\Delta_{\pi}/\Omega_{\mathrm{\pi}} = 1$ and $1.5$. Outside this detuning range, the emergence of the first bound state at lower field strengths restricts the optimal Rabi frequency back below 20~MHz. In contrast, the heavy, strongly dipolar {CsAg} molecule requires much lower field strengths to achieve extreme loss suppression ($\sim 10^{-17}$ cm$^{3}$/s). By avoiding the upper cap entirely, CsAg outperforms NaCs by two orders of magnitude. {However, the favorable mass and dipole scaling introduce an important nuance visible in the CsAg curves: the width of the accessible detuning range is narrower for CsAg than for NaCs. Just as a larger dipole moment reduces the required Rabi frequency to reach the universal optimum, it simultaneously pushes the threshold for the appearance of field-linked bound states to much lower field strengths. As a result, the optimal Rabi frequency for CsAg drops below our practical lower bound of 1~MHz sooner than it does for NaCs, effectively restricting the width of the operational window. Rather than a fundamental trade-off, this indicates that for heavy, dipolar species, the \textit{lower} limit of the available field strengths acts as the primary bottleneck for the usable part of the parameter space. Nevertheless, within this narrower window, they unequivocally provide the deepest and most robust loss suppression.}

In the same panel, we also show the results obtained when accounting for the ellipticity of the microwave fields (dashed lines). {As detailed in Sec.~\ref{subsec:ellipticity}, we assume realistic imperfect polarizations of microwave fields: an ellipticity of $0.5^{\circ}$ for the $\sigma^{+}$ field and transverse components up to $4^{\circ}$ for the $\pi$ field, which are achieved in experimental setups \cite{Schindewolf_2022,Bigagli_2023,Bigagli_2024,Chen_2023,Lin2023}.}
While non-zero ellipticity increases the absolute loss rate by up to a factor of 3, it does not qualitatively alter the shape of the loss curves or the underlying conclusions regarding species scaling.

Figure~\ref{fig:1d_cuts} illustrates also the tunability of the effective dipolar interactions. The solid green line tracks the effective dipolar length, expressed in universal units ($a^{\mathrm{eff}}_{dd} / a_{dd}$). We note that the deep minimum of the loss curve in the left panel of Fig.~\ref{fig:1d_cuts} does not coincide with the compensation point ($a^{\mathrm{eff}}_{dd} = 0$), but rather occurs at small, negative effective dipolar lengths. To highlight the range of tunability of dipolar interactions, the cross ($\times$) marker identifies the largest negative effective dipolar length achievable {with both molecules} while keeping the loss rate below the $10^{-13}$ cm$^{3}$/s threshold. 
The right column [panels (b) and (d)] complements this by tracking a diagonal cut ({$\Delta_{\sigma}/\Omega_{\sigma} = \alpha\,\Delta_{\pi}/\Omega_{\pi} + \beta$}) through the balanced field configuration ($\Omega_{\sigma} / \Omega_{\pi} = 1$, shown in the center panels of Fig.~\ref{fig:loss_dipole}) {for $\alpha = 0.68, \, \beta = -0.072$}. This specific trajectory passes the complementary extreme: the plus ($+$) markers designate the largest positive effective dipolar length that can be reached without violating the same $10^{-13}$ cm$^{3}$/s loss threshold. The required physical field strengths for this trajectory are shown in panel (d). Because this cut operates close to the compensation point, where the optimal Rabi frequencies are large, NaCs is restricted by the 20~MHz cap almost immediately (starting at$\Delta_{\pi} /\Omega_{\pi} \approx 0.36$) and remains capped for the rest of the scan.  {Consistent with the $\pi$-dominated case, we again observe that CsAg achieves deeper loss suppression---here by an order of magnitude---but over a narrower range of detunings, as its optimal Rabi frequency drops below the 1~MHz threshold sooner than for NaCs. While this narrower window visually constrains the maximum fractional dipole moment (the $+$ marker), this is merely an artifact of dimensionless units: the permanent dipole moment of CsAg overwhelmingly compensates for the smaller fractional range, preserving vast tunability as detailed in Sec.~\ref{subsec:tunability}.}

\subsection{Prospects for evaporative cooling}
\label{subsec:evaporation}

\begin{figure*}
    \centering
    \includegraphics[width=0.7\linewidth]{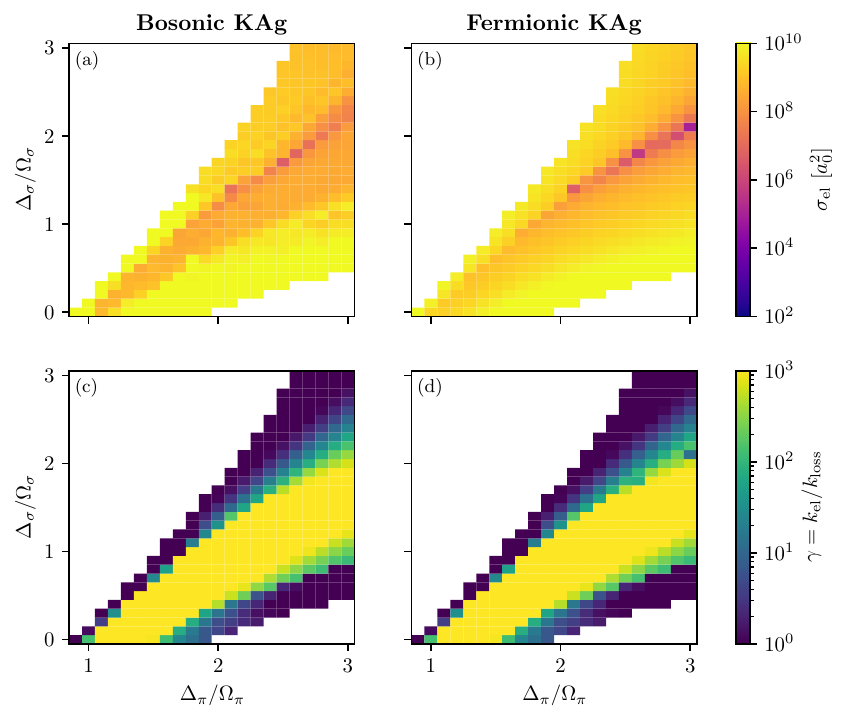}
    \caption{Prospects for evaporative cooling of bosonic and fermionic KAg molecules at $T=1\,\mu\mathrm{K}$ using double microwave shielding at the $\Omega_{\sigma}/\Omega_{\pi}=1/4$ field configuration. (a, b) elastic cross-section, $\sigma_{\mathrm{el}}$. (c, d) The ratio of the elastic-to-inelastic rate coefficients, $\gamma$. {The color scale in (c, d) is saturated at a maximum practical value of $10^{3}$, set by the hydrodynamic thermalization rate and the one-body lifetime.}}
    \label{fig:elastic_xs_gamma}
\end{figure*}

So far, we have focused exclusively on minimizing two-body loss. However, from the standpoint of creating quantum degenerate gases via evaporative cooling, the critical figure of merit is the ratio of elastic to inelastic collision rates, $\gamma = k_{\mathrm{el}} / k_{\mathrm{loss}}$. To sustain efficient evaporation, this ratio must typically exceed 100~{\cite{Monroe_1993, Ketterle_1996}}. This requirement introduces an interesting trade-off for double-microwave shielding: while perfect compensation of the effective dipole moment {reduces} losses, it simultaneously damps the long-range dipolar collisions necessary for thermalization.
This issue is of particular importance for fermionic molecules, where identical particles in the same internal state are protected from short-range contact interactions by the $p$-wave centrifugal barrier, leaving long-range dipolar scattering as the sole mechanism for thermalization.

To explore this trade-off, we map the parameter space to identify the optimal conditions for evaporative cooling, selecting KAg as a representative species of interest that can be formed as both bosonic and fermionic isotopes. For bosonic KAg, we compute the elastic and inelastic rates directly from Eq.~\eqref{eq:loss_rate_again} and Eq.~\eqref{eq:elastic_rate}. For identical fermions, the $s$-wave scattering length vanishes. The purely dipolar elastic rate coefficient for fermions is given by~\cite{Bohn_2009}
\begin{equation}
    k_{\mathrm{el}}^{(f)} = \frac{32 \pi}{15}\langle v \rangle \,{a_{dd}^{\mathrm{eff}}}^{2},
\end{equation}
which we compute using effective dipolar lengths extracted from our scans of the parameter space. For the fermionic loss rate, we adopt the computed bosonic $s$-wave loss rate [Eq.~\eqref{eq:loss_rate}] as a conservative upper bound for the actual loss rates, which are further suppressed by the $p$-wave barrier.
The true fermionic loss rates will be lower, and the resulting $\gamma$ ratios higher, than our estimates. We evaluate these rates at a fixed temperature of 1 $\mu$K. 

To ensure the calculated ratios reflect realistic experimental conditions, we impose bounds on the {relevant timescales.} For fast elastic collisions, {the thermalization rate is ultimately limited by the trap frequencies (typically $\sim100\,\mathrm{Hz}$), as the system enters the hydrodynamic regime~\cite{Ma_2003, Schindewolf_2022}.
Assuming the density of the sample of $10^{12}$~cm$^{-3}$, we limit $k_{\mathrm{el}}$ at a $10^{-10}$~cm$^3$/s since even faster elastic collisions do not improve thermalization.
The practical limit for loss suppression is set by the one-body lifetime of the sample, set, for example, by off-resonant scattering of the
trap light (typically $\sim 10\,\mathrm{s}$). 
Therefore, we limit $k_{\mathrm{loss}}$ from below at $10^{-13}$~cm$^{3}$/s.
These two limits effectively cap the useful ratio of elastic to inelastic collisions at $\gamma_{\mathrm{max}} = 10^{3}$, the ratio of the thermalization rate in the hydrodynamic regime and the one-body loss rate. Any theoretical ratio exceeding this value provides no further practical benefit for evaporative cooling.}

Figure~\ref{fig:elastic_xs_gamma} presents the elastic cross-sections, $\sigma_{\mathrm{el}}$, and the resulting ratio of elastic-to-inelastic rate coefficients, $\gamma$ computed at $1\,\mu \mathrm{K}$ for both bosonic and fermionic KAg. We focus on the $\pi$-dominated field configuration ($\Omega_{\sigma}/\Omega_{\pi} = 1/4$), which we previously identified as providing the broadest parameter space for extreme loss suppression. The results show striking similarities between both the bosonic and fermionic systems. For a heavy, strongly dipolar molecule like KAg, where the achievable effective dipolar lengths are on the order of $10^{4}\,a_{0}$, the long-range dipolar scattering dominates collision dynamics, driving the elastic cross-sections ({Fig.~\ref{fig:elastic_xs_gamma} (a, b)}) well into the $10^{7}-10^{9}\,a_{0}^{2}$ range. Because these values {easily saturate the hydrodynamic thermalization limit, the corresponding $\gamma$ ratio (Fig.~\ref{fig:elastic_xs_gamma} (c, d)}) is effectively dominated by the loss rate. Wherever the two-body loss is sufficiently suppressed to exceed the one-body lifetime, $\gamma$ saturates at its maximum practical value of $10^{3}$, {set by the hydrodynamic limit and the one-body lifetime}. As the figure shows, this condition is easily met, yielding a vast, highly favorable region for both isotopes.

The most prominent difference between the {bosonic and fermionic species} is seen along the narrow diagonal stripe at high detuning ratios. This stripe tracks the vicinity of the compensation point, where the two fields effectively cancel the long-range dipolar interaction. For bosons, $k_{\mathrm{el}}$ is then determined by the $s$-wave scattering length contribution.
For fermions, the compensated dipolar interaction causes the elastic cross-sections to drop to zero along this line. However, this narrow feature is of little practical concern: the optimal parameter regime for evaporative cooling does not coincide with the compensation stripe. Instead, the largest values of $\gamma$ are achieved by maintaining a finite effective dipole moment.  {Ultimately}, Fig.~\ref{fig:elastic_xs_gamma} clearly shows a broad range of parameters where both isotopes simultaneously exhibit large elastic rates and deeply suppressed losses, confirming that double microwave shielding provides an effective platform for evaporative cooling for both bosonic and fermionic molecules.

{We note that if the limits of the timescales are lifted, the \textit{unconstrained} $\gamma$ ratio in these optimal regimes can reach as high as $10^{7}$. 
In other words, in practice, realizable $\gamma$ ratios may be limited primarily by technical aspects of the experiment rather than the quality of shielding.
Access to this enormous underlying $\gamma$ buffer is particularly important for fermionic molecules. As the Fermi gas approaches deep degeneracy, Pauli blocking severely suppresses the effective elastic collision rate, as molecules can only scatter into a decreasing number of unoccupied states near the Fermi surface~\cite{DeMarco2001,Aikawa2014,Aikawa2014b}. The immense unconstrained $\gamma$ therefore provides a crucial margin to sustain thermalization much deeper into the quantum degenerate regime. Specifically, it could enable slow evaporation of low-density samples, potentially in a box trap, provided that one-body lifetimes are sufficiently long.}

\subsection{Tunability of dipolar interactions}
\label{subsec:tunability}
\begin{figure*}
    \centering
    \includegraphics[width=0.8\linewidth]{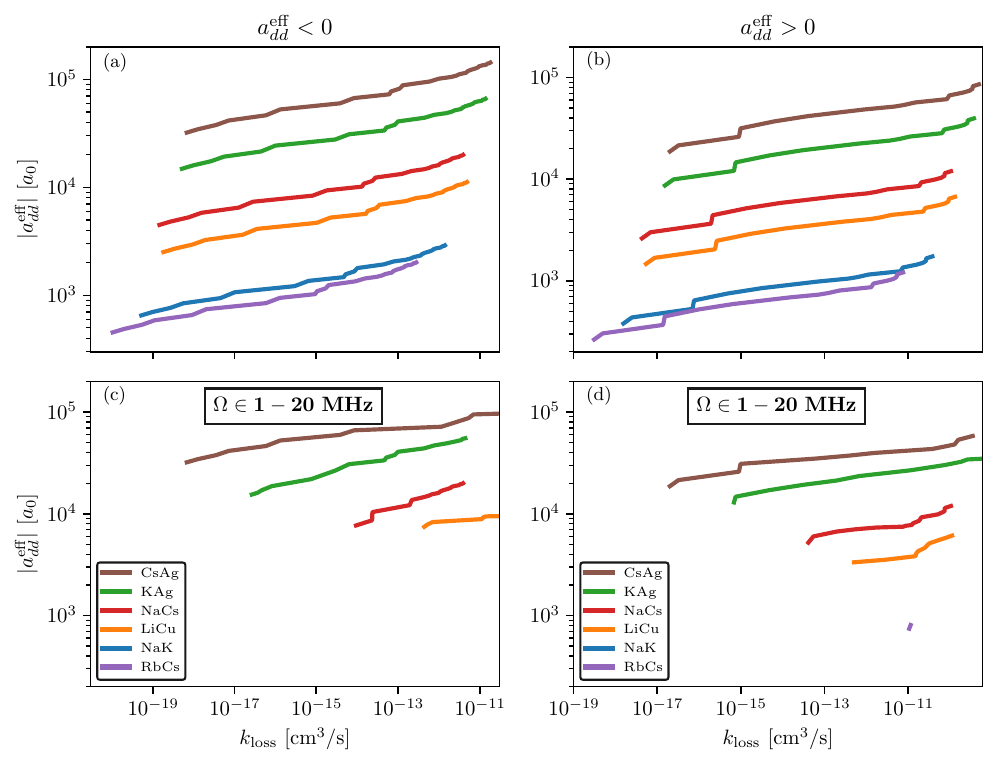}
    \caption{(a, b) the largest negative (left) and positive (right) dipolar length achievable at a given loss rate for selected molecules (no constraints on $\Omega$). (c, d) same as top, but $\Omega_{\sigma}$ and $\Omega_{\pi}$ are now limited to $1-20$~MHz.
    }
    \label{fig:dipole_binning}
\end{figure*}

{In the previous sections, we mapped the double-microwave parameter space to identify the optimal conditions for loss suppression and efficient evaporative cooling.} In many experiments, however, the goal is not necessarily to fully eliminate collisional two-body loss, but rather to suppress it sufficiently to conduct the experiment at hand, while simultaneously using the microwave fields to induce and tune dipolar interactions. {As established earlier, suppressing the two-body loss rate to the level of $k_{\mathrm{loss}} \leq 10^{-13}$~cm$^{3}$/s is typically sufficient, as the collisional lifetime then exceeds typical one-body lifetimes. Rather than treating this value as a rigid cutoff, we now explore the continuous trade-off between the maximum achievable interaction strength and the tolerated two-body loss rate.}

Figure~\ref{fig:dipole_binning} illustrates this tunability for various bialkali and coinage-metal systems. We plot the absolute magnitude of the achievable effective dipolar length, $|a^{\mathrm{eff}}_{dd}|$, as a function of the allowed two-body loss rate for both the anti-dipolar ($a_{dd}^{\mathrm{eff}} < 0$, left panels) and dipolar ($a_{dd}^{\mathrm{eff}} > 0$) interactions. The top panel presents the unconstrained case, where Rabi frequencies are allowed to take any value required to reach a specific effective interaction strength. {Here,} we observe a clear manifestation of the universality in double microwave shielding: the curves for different species share an identical shape, shifted along the horizontal axis according to $d^{2}$, and along the vertical axis according to $\mu d^{2}$. {At the absolute peak of these unconstrained curves, the effective interaction corresponds to an induced dipole moment ($d_{\mathrm{eff}} \propto \sqrt{|a_{dd}^{\mathrm{eff}}|}$) reaching roughly 20\% of the molecule's bare permanent dipole moment.}

To assess realistic experimental capabilities, we apply physical constraints to the Rabi frequencies, restricting them to the range $1 \, \mathrm{MHz} \le \Omega \le 20 \, \mathrm{MHz}$ (Fig.~\ref{fig:dipole_binning}, bottom panel). Imposing these limits {truncates the achievable tunability range}. 
For species with smaller bare dipoles like RbCs and NaK, the Rabi frequencies required to access the low-loss high-tunability regime lie far above 20~MHz. Consequently, with the feasibility window defined above, these species offer no tunability. {For} intermediate species like NaCs and LiCu, {the field strength constraint visibly truncates their curves in the lower panels. However, they can still generate large interaction strengths, easily reaching dipolar lengths on the order of $10^{4}\,a_0$.}
In contrast, strongly dipolar species such as CsAg and KAg retain almost their full range of tunability. Because their optimal operating points naturally fall within the few-MHz regime, they can achieve large absolute dipolar lengths ($\sim 10^{5}\,a_{0}$) at two-body loss rates that remain orders of magnitude lower than those of the lighter molecules. This analysis confirms that for experiments constrained to moderate field strengths, strongly dipolar molecules offer the most robust platform for quantum simulation, providing significant tunability while keeping two-body loss rates at manageable levels.

{We note that these large effective length scales can easily approach or exceed typical trap confinement lengths ($\sim 10^{4}\,a_{0}$~\cite{Bigagli_2024,Shi_2026}). Under such conditions, collision dynamics become influenced by the trap geometry, which opens the possibility of accessing confinement-induced resonances \cite{Olshanii1998,Haller2010,dunjko2011confinement,capecchi2023}. These resonances occur when the effective interaction length scale becomes comparable to the harmonic oscillator length of the trap, leading to a strong divergence in the low-dimensional interaction strength. Reaching this regime with shielded polar molecules would provide a compelling pathway to explore stable, strongly interacting molecular dipolar gases in reduced dimensions~\cite{Chomaz_2022}, and potentially enable the formation of confinement-induced bi-molecular bound states (tetramers), analogous to the geometrically associated atomic dimers observed in ultracold atomic gases~\cite{Sala2013}.}

\subsection{Survey of candidate polar molecules}
\label{subsec:molecules}
Finally, we extend our analysis beyond specific molecular examples discussed so far to provide a comprehensive outlook for the field. By exploiting the universality of the scattering problem, we map our results {across a} continuous range of masses and dipole moments. This allows us to identify which candidate molecules offer the optimal combination of low loss and high tunability under realistic experimental constraints. 

\begin{figure*}[!ht]
    \centering
    \includegraphics[width=0.49\linewidth]{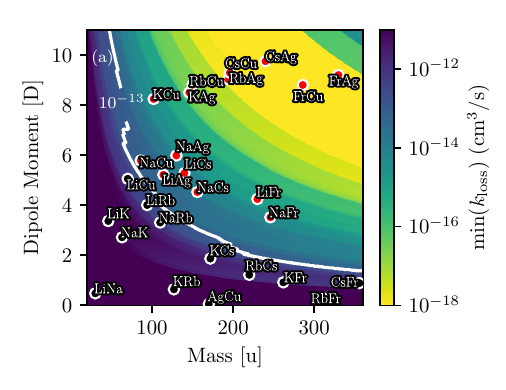}
    \includegraphics[width=0.49\linewidth]{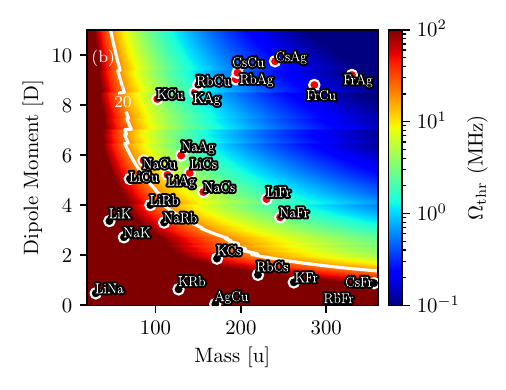}
    \caption{{(a)} The absolute lowest loss rate achievable with $\Omega$ ranging between 1 and 20~MHz. {(b)} The minimum Rabi frequency, ${\Omega_{\mathrm{thr}}}$, needed to {suppress the two-body} loss rate below $10^{-13}$~cm$^{3}$/s.}
    \label{fig:global_minimum_loss_rabi}
\end{figure*}

Figure~\ref{fig:global_minimum_loss_rabi} (left panel) presents {the absolute minimum} loss rate achievable with Rabi frequencies restricted to the feasible range of $1-20$~MHz. We {continuously} scan the {mass-dipole moment plane}, marking the positions of relevant alkali dimers, as well as silver- and copper-bearing species. A contour line indicates the threshold of $10^{-13}$~cm$^3$/s: species located above this line are capable of suppressing two-body losses to $10^{-13}$~cm$^3$/s or less using only moderate fields. 

The trends observed here generalize the conclusions drawn from Fig.~\ref{fig:loss_binning}. When non-universal constraints (constraints in absolute units on dimensionful quantities, e.g., $\Omega \in 1-20$~MHz) are imposed, heavy, strongly dipolar species consistently outperform their lighter, less-polar counterparts. {This conclusion differs from that of single-microwave shielding, in which we found that weakly dipolar molecules are favored. The reason for} this reversal {lies in the distinct physical mechanisms governing the two shielding techniques.}  In single microwave shielding, the maximum achievable dimensionless Rabi frequency ($\tilde{\Omega}$) is strictly capped by the appearance of a field-linked bound state. Because of this limit, the dimensionless loss $\tilde{\beta}$ cannot be arbitrarily suppressed and remains at the $ \tilde{\beta } \geq 10^{-5}$ level. The absolute physical loss ($k\propto d^{2} \tilde{\beta}$) becomes then dominated by the $d^{2}$ prefactor, favoring weakly dipolar species. In double microwave shielding, {however}, the $\pi$-field compensates the attractive well, effectively removing this bound-state bottleneck. The {unconstrained theoretical} global minimum now lies at an extremely large {dimensionless} Rabi frequency ($\tilde{\Omega}^{\mathrm{tot}} \sim 10^{10}$).
At this {unconstrained} minimum, $\tilde{\beta}$ drops by eight orders of magnitude, a suppression that easily overwhelms the $d^{2}$ prefactor penalty. 

However, to access this minimum using only experimentally feasible fields ($\Omega=1-20$~MHz), the scaling relation $\Omega \propto \tilde{\Omega} \mu^{-3} d^{-4}$ becomes the decisive factor. For heavy, ultra-polar molecules, a moderate physical field ($\Omega$) corresponds to an extremely large dimensionless field ($\tilde{\Omega}$), enabling access to the extreme loss suppression regime. Conversely, for light, less dipolar molecules, the same physical field corresponds to a much lower $\tilde{\Omega}$. This is illustrated in the left panel of Fig.~\ref{fig:global_minimum_loss_rabi}: heavy, ultra-polar species formed with Ag and Cu (e.g., CsAg, RbAg) can achieve loss rates as low as $10^{-18}$~cm$^3$/s with experimentally feasible fields -- orders of magnitude below the {$10^{-13}$ cm$^{3}$/s threshold at which two-body collisions become negligible compared to typical one-body lifetimes.} Meanwhile, a broad class of lighter species with intermediate dipole moments ($4-6$ D) also comfortably reach the sub-$10^{-13}$~cm$^3$/s regime.

{The results presented in Fig.~\ref{fig:global_minimum_loss_rabi}(a) illustrate the fundamental molecular properties that govern the efficiency of double microwave shielding. 
The decisive factor is not merely the molecule's dipolar length ($a_{dd} \propto \mu d^{2}$), but more precisely the characteristic dipolar energy scale (${E_{dd} \propto 1/(\mu a_{dd}^{2})}$), which dictates the conversion between dimensionless and physical fields, on which we impose constraints. Because of the strong, fourth-power scaling of $E_{dd}$ with $d$, a large permanent dipole moment is the most critical asset. However, if two candidate molecules were to possess comparable dipolar lengths, the heavier species would be advantageous: its larger mass further reduces $E_{dd}$, providing a much larger dimensionless Rabi frequency $\tilde{\Omega}$ for the same applied microwave power. Consequently, the color contours indicating different levels of loss suppression in Fig.~\ref{fig:global_minimum_loss_rabi}(a) do not follow paths of constant $a_{dd}$, but instead closely trace lines of constant maximum dimensionless Rabi frequency, $\tilde{\Omega} \propto \mu^{3}d^{4}$.}

To view this from a complementary perspective, the right panel of Fig.~\ref{fig:global_minimum_loss_rabi} plots the minimum physical Rabi frequency required to suppress losses below the $10^{-13}$~cm$^3$/s threshold. The color map mirrors the previous trend: heavy, strongly dipolar molecules (CsAg, CsCu, RbAg) can reach this suppression level with fields well below $1$~MHz. Markers indicate feasibility: red markers denote molecules where this suppression is achievable within the {$1-20$~MHz} limit, which includes intermediate species like LiRb and NaRb, while black dots denote species like LiCu or RbCs, where the required fields far exceed experimentally achievable field strengths. {Furthermore, since experimental limitations are typically set by the maximum applied electric field $E$, and the Rabi frequency scales as $\Omega \propto d \cdot E$, achieving these threshold Rabi frequencies for weakly dipolar molecules requires proportionally stronger electric fields, making the practical advantage of highly polar species even more pronounced.}

\begin{figure*}[!ht]
    \centering
    \includegraphics[width=0.49\linewidth]{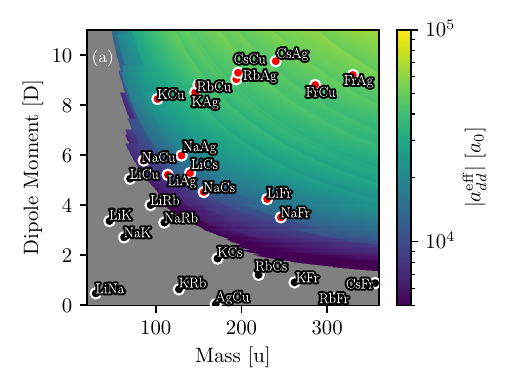}
    \includegraphics[width=0.49\linewidth]{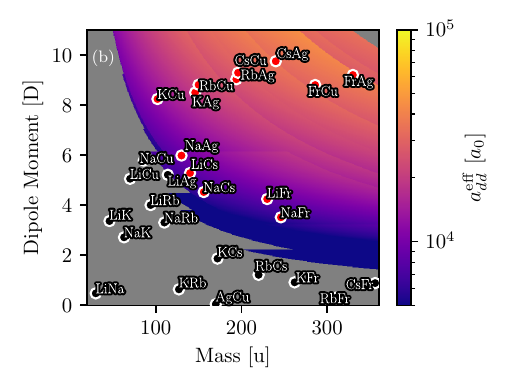}
    \caption{The largest negative {(a)} and positive {(b)} dipolar lengths achievable with $\Omega$ ranging between 1 and 20~MHz and loss rates below $10^{-13}$~cm$^{3}$/s. {The grey regions in the bottom-left corners correspond to light, weakly dipolar species for which the 1--20~MHz field strength is insufficient to suppress two-body losses below this threshold, leaving them with no practical tunability range.}}
    \label{fig:dipole_tunability}
\end{figure*}

Finally, we address the tunability of the effective dipolar interaction expressed as the maximum positive and negative effective dipolar length, $a_{dd}^{\mathrm{eff}}$, restricted to configurations that satisfy both the loss suppression criterion ($k < 10^{-13}$~cm$^3$/s) and the experimental feasibility criterion ($\Omega = 1- 20$~MHz). Figure~\ref{fig:dipole_tunability} displays the maximum achievable negative (left panel) and positive (right panel) effective dipolar lengths across the mass-dipole plane.

The results confirm the trend established in Fig.~\ref{fig:global_minimum_loss_rabi}: the combination of low mass and small dipole moment precludes any meaningful tunability within the $20$~MHz limit. Molecules such as LiCu, NaRb, and RbCs effectively have zero tunable range under these constraints. Moving to {intermediate masses and dipole moments}, species like NaCs provide a substantial tuning range of approximately $\pm 10^{4} \, a_0$. The broadest tunability ranges are found in the heavy, ultra-polar species (RbAg, CsCu, CsAg), where the effective interaction can be tuned over a range reaching $\pm 10^{5} \, a_0$. {In the absence of constraints, the maximum tunable effective dipolar length scales trivially with the bare dipolar length ($a^{\mathrm{eff}}_{dd} \propto a_{dd}$). As we move upward from moderately dipolar species (e.g., NaCs) to highly dipolar ones (e.g., CsAg), the achievable tunability grows precisely according to this universal expectation (see also Fig.~\ref{fig:dipole_binning}). However, this straightforward linear scaling breaks down when moving in the opposite direction towards less polar species. Because lighter, weakly interacting molecules are severely restricted by the limit on physical field strengths, long before they can reach their theoretical optimum, their practical tunability diminishes much faster than their bare $a_{dd}$ would suggest. Ultimately, just as with loss suppression, the decisive parameter combination determining practical tunability under realistic experimental constraints is not simply $a_{dd}$, but the $\mu^{3} d^{4}$ scaling that governs the maximum achievable dimensionless field.}
 This demonstrates that for future experiments aiming to explore strong, tunable dipolar physics with long lifetimes, ultra-polar molecules offer a substantial advantage.

\section{Conclusions}
\label{sec:conclusions}
In this work, we provided a comprehensive theoretical investigation of double microwave shielding for ultracold polar molecules. By leveraging the universality of this problem, we systematically mapped the four-dimensional microwave parameter space to identify the optimal conditions for minimizing two-body collisional loss and maximizing the tunability of dipolar interactions, while simultaneously ensuring the absence of field-linked bound states that lead to three-body recombination.

Unlike single microwave shielding, where the appearance of field-linked bound states strictly caps the achievable loss suppression, we demonstrated that the addition of the $\pi$-polarized field effectively removes this bottleneck. By reducing the depth of the long-range attractive well created by the $\sigma^{+}$ field, double microwave shielding allows achieving a theoretical global optimum where two-body loss rates drop by eight orders of magnitude compared to the lowest limit achievable in the single-field case.

Crucially, we showed that translating this universal theoretical optimum into an experimentally feasible range of microwave parameters fundamentally alters the hierarchy of candidate molecules. The required physical microwave field strength is determined by the characteristic dipolar energy, which introduces a $\mu^{-3} d^{-4}$ scaling with the system's reduced mass and permanent dipole moment. When constrained to realistic field strengths quantified by the Rabi frequencies in the range $1-20$~MHz, this scaling penalizes light, weakly dipolar molecules, severely limiting their accessible loss suppression and tunability. In stark contrast, for ultra-polar molecules (such as CsAg, KAg, and RbAg), moderate microwave fields are sufficient to suppress absolute two-body loss rates to the $10^{-18}$~cm$^{3}$/s level, ensuring two-body collisional lifetimes that far exceed the typical one-body lifetimes in recent experiments.

Furthermore, we established that these deeply shielded regimes are highly favorable for the creation of quantum degenerate gases. Both bosonic and fermionic species exhibit large elastic-to-inelastic collision ratios ($\gamma \sim 10^3$), easily satisfying the requirements for efficient evaporative cooling. Concurrently, the double-field scheme allows the effective dipolar lengths to be tuned over an immense range (up to $\pm 10^5\, a_0$), without compromising the collisional stability of the samples.

Ultimately, our broad survey across a wide range of molecular species demonstrates that the combination of heavy mass and large permanent dipole moment is the decisive factor for successful double microwave shielding. This establishes heavy, strongly dipolar species like the coinage-metal dimers as the most robust and versatile platforms for future experiments with ultracold molecules.

\section{Acknowledgment}

The research was funded by the European Union (Project No.~101269084, HORIZON-MSCA-2025-PF, 2STICKY). The views and opinions expressed are, however, those of the authors only and do not necessarily reflect those of the European Union or the European Research Executive Agency. Neither
the European Union nor the granting authority can be held responsible for them.
T.K.~is supported by NWO VIDI (grant ID 10.61686/AKJWK33335). S.W.~acknowledges support from the NSF (Award No.~2409747), AFOSR (Award No.~FA9550-25-1-0048), and the Gordon and Betty Moore Foundation (Award No.~GBMF12340).
I.S.~ acknowledges support form the Ernest Kempton Adams Fund and the University of Virginia.
This work was performed in part at the Aspen Center for Physics, which is supported by National Science Foundation grant PHY-2210452.

\bibliography{bibliography}
\end{document}